\documentclass[
prd,preprint
,superscriptaddress,noshowpacs,nofootinbib
,tightenlines
]{revtex4}
\usepackage{mathrsfs}
\usepackage{latexsym,bm}
\usepackage{graphicx}
\usepackage{indentfirst}
\usepackage{slashed}
\usepackage{amssymb}
\usepackage{amsmath}
\usepackage{bbm}
\usepackage{color}
\usepackage[dvipsnames]{xcolor}
\usepackage{epsfig}
\usepackage[titletoc]{appendix}
\usepackage{multirow}%
\usepackage{rotating}
\usepackage{epstopdf}
\usepackage{extarrows}
\usepackage{tabularx}
\usepackage{soul} 
\usepackage{xcolor}
\usepackage{lipsum}
\usepackage[colorlinks=true,allcolors=BlueViolet,hyperfootnotes=false]{hyperref}

\usepackage[utf8]{inputenc}

\newcommand{\be}{\begin{equation}} \newcommand{\ee}{\end{equation}}
\newcommand{\ba}{\begin{array}{c}} \newcommand{\ea}{\end{array}}
\newcommand{\bea}{\begin{eqnarray}} \newcommand{\eea}{\end{eqnarray}}

\newcommand{\no}{\nonumber\\}
\newcommand{\MOA}{{\cal O}^A}
\newcommand{\MOV}{{\cal O}^V}
\newcommand{\MOACES}{\tilde{{\cal O}}^A}
\newcommand{\MOVCES}{\tilde{{\cal O}}^V}
\newcommand{\gA}{g}
\newcommand{\mn}{m}
\newcommand{\mnuc}{m_N}
\newcommand{\mpir}{M_\pi}
\newcommand{\mdel}{m_\Delta}
\newcommand{\F}{F_\pi}
\newcommand{\g}{g_A}
\newcommand{\h}{h_A}

\newcommand{\Aone}{\tilde{A}_1}
\newcommand{\Atwo}{\tilde{A}_2}
\newcommand{\Athr}{\tilde{A}_3}
\newcommand{\Afou}{\tilde{A}_4}
\newcommand{\Afiv}{\tilde{A}_5}
\newcommand{\Asix}{\tilde{A}_6}
\newcommand{\Asev}{\tilde{A}_7}
\newcommand{\Aeig}{\tilde{A}_8}
\newcommand{\Vone}{\tilde{V}_1}
\newcommand{\Vtwo}{\tilde{V}_2}
\newcommand{\Vthr}{\tilde{V}_3}
\newcommand{\Vfou}{\tilde{V}_4}
\newcommand{\Vfiv}{\tilde{V}_5}
\newcommand{\Vsix}{\tilde{V}_6}

\newcommand{\Dst}{\Delta_{s_2}}
\newcommand{\Sst}{\Sigma_{s_2}}
\newcommand{\Dstpr}{\Delta_{s_2}^\prime}
\newcommand{\Sstpr}{\Sigma_{s_2}^\prime}
\newcommand{\Du}{\Delta_{u}}
\newcommand{\Su}{\Sigma_{u}}
\newcommand{\Dupr}{\Delta_{u}^\prime}
\newcommand{\Supr}{\Sigma_{u}^\prime}

\newcommand{\LFAoBar}{\bar{A}_0}
\newcommand{\LFBoBar}{\bar{B}_0}
\newcommand{\LFAo}{{A}_0}
\newcommand{\LFBo}{{B}_0}

\newcommand{\qdotq}{|\vec{{q}}|^2}
\newcommand{\kdotk}{|\vec{{k}}|^2}
\newcommand{\absq}{|\vec{{q}}|}
\newcommand{\absk}{|\vec{{k}}|}
\newcommand{\kdotq}{\vec{{k}}\cdot\vec{{q}}}

\allowdisplaybreaks

\begin{document}

\title{\LARGE  Weak pion production off the nucleon \\
in covariant chiral perturbation theory
}
\author{De-Liang Yao}
\email{Deliang.Yao@ific.uv.es}
\affiliation{Departamento de F\'{\i}sica Te\'orica and Instituto de F\'{\i}sica Corpuscular (IFIC), Centro Mixto CSIC-UV, Institutos de Investigaci\'on de Paterna,
E-46071, Valencia, Spain}
\author{Luis Alvarez-Ruso}
\email{Luis.Alvarez@ific.uv.es}
\affiliation{Departamento de F\'{\i}sica Te\'orica and Instituto de F\'{\i}sica Corpuscular (IFIC), Centro Mixto CSIC-UV, Institutos de Investigaci\'on de Paterna,
E-46071, Valencia, Spain}
\author{Astrid N.~Hiller Blin}
\email{Hillerbl@uni-mainz.de}
\affiliation{Institut f\"ur Kernphysik \& PRISMA Cluster of Excellence, Johannes Gutenberg Universit\"at, D-55099 Mainz, Germany}
\author{M. J.~Vicente Vacas}
\email{Manuel.J.Vicente@ific.uv.es}
\affiliation{Departamento de F\'{\i}sica Te\'orica and Instituto de F\'{\i}sica Corpuscular (IFIC), Centro Mixto CSIC-UV, Institutos de Investigaci\'on de Paterna,
E-46071, Valencia, Spain}

\date{\today}
\begin{abstract}
  Weak pion production off the nucleon at low energies has been systematically investigated  in manifestly relativistic baryon chiral perturbation theory with  explicit inclusion of the $\Delta$(1232) resonance. Most of the involved low-energy constants have been previously determined in other processes such as pion-nucleon elastic scattering and electromagnetic pion production off the nucleon. For numerical estimates, the few remaining constants are set to be of natural size. As a result, the total cross sections for single pion production on neutrons and protons, induced either by neutrino or antineutrino, are predicted. Our results are consistent with the scarce existing experimental data except in the $\nu_\mu n\to \mu^-n\pi^+$ channel, where higher-order contributions might still be significant. The $\Delta$ resonance mechanisms lead to sizeable contributions in all channels, especially in $\nu_\mu p\to \mu^- p\pi^+$, even though the considered energies are close to the production threshold. The present study provides a well founded low-energy benchmark for phenomenological models aimed at the description of weak pion production processes in the broad kinematic range of interest for current and future neutrino-oscillation experiments. 
\end{abstract}
\maketitle


\section{Introduction}

Neutrino interactions with matter are at the heart of many relevant phenomena in astrophysics, nuclear and particle physics. Among them, neutrino oscillations have revealed that neutrinos are massive, providing evidence of physics beyond the Standard Model. Precision studies of neutrino-oscillation parameters demand a good understanding and accurate modeling of neutrino interactions with nucleons and nuclei~\cite{Formaggio:2013kya,Alvarez-Ruso:2014bla,Alvarez-Ruso:2017oui}. In this context, weak pion production has been actively investigated.   

Single-pion production amounts to one of the leading contributions to the inclusive (anti)neutrino-nucleus
cross section in the energy range of interest for ongoing and future oscillation experiments. As such, it can be part of the signal or a background that should be precisely constrained. Single charged pion production in charged-current interactions is a source of events that can be misidentified as quasielastic [$\nu_l(\bar{\nu}_l) N \rightarrow l^\mp N'$] ones if the pion is not identified, introducing a bias in the kinematic neutrino energy reconstruction~\footnote{Neutrino fluxes are not monochromatic. Therefore, the neutrino energy, on which oscillation probabilities depend, is not known on an event-by-event basis but can be approximately reconstructed from the final-lepton kinematics in quasielastic events.}. Furthermore neutral-current $\pi^0$ production events in Cherenkov detectors contribute to the electron-like background in $\nu_\mu \to \nu_e$ measurements. In spite of the progress, 20-30\% errors are currently taken for single-pion production in oscillation analyses due to conflicts between data sets and models~\cite{Katori:2016yel}. 

It was early acknowledged that, at low and intermediate energies, weak pion production should proceed predominantly through the excitation of the $\Delta(1232)3/2^+$ resonance (see Ref.~\cite{LlewellynSmith:1971uhs} and references therein). Isobar models accounting for heavier nucleon resonances were subsequently developed~\cite{Fogli:1979cz,Rein:1980wg}. The nucleon-to-resonance transitions were  parametrized in terms of real form factors obtained from quark models~\cite{Rein:1980wg,Hemmert:1994ky,Golli:2002wy,BarquillaCano:2007yk} or phenomenology. In the later case, owing to the symmetry of the conserved vector current under isospin rotations, vector transition form factors can be related to electromagnetic ones extracted from electron scattering data while the partial conservation of the axial current (PCAC) allows to derive the off-diagonal Goldberger-Treiman (GT) relation for the leading axial couplings~\cite{Fogli:1979cz,Lalakulich:2006sw,Leitner:2008ue,Alam:2015gaa}. Additional, but rather limited, information on the transition axial form factors can be obtained from available weak pion-production bubble-chamber data on hydrogen and deuterium~\cite{Schreiner:1973ka,Hernandez:2010bx,Graczyk:2014dpa}. Non-resonant mechanisms were added to the resonant ones in Refs.~\cite{Bijtebier:1970ku,Alevizos:1977xf,Fogli:1979cz}, and further extended to fulfill chiral symmetry constraints at threshold in Ref.~\cite{Hernandez:2007qq}. In these studies, the range of applicability of the Born terms is expanded by the introduction of form factors. In the approach of Ref.~\cite{Hernandez:2007qq}  (denoted as HNV from now on), a good agreement with bubble chamber data was achieved at the price of introducing tensions in the value of the leading $N-\Delta(1232)$ axial coupling, $C_5^A$ in the notation of Ref.~\cite{LlewellynSmith:1971uhs}, with respect to the GT value at a 2$\sigma$ level~\cite{Hernandez:2010bx}\footnote{Deviations from the $N-\Delta(1232)$ off-diagonal GT relation are expected only at the few-$\%$ level, as they arise from chiral symmetry breaking. Systematic studies of the corrections to this GT relation using chiral perturbation theory have been reported in Refs.~\cite{Geng:2008bm,Procura:2008ze}.}. The two values could be reconciled by imposing Watson's theorem in the dominant partial wave~\cite{Alvarez-Ruso:2015eva}. The importance of a consistent treatment of the $\Delta(1232)$ was stressed in Refs.~\cite{Barbero:2008zza,Barbero:2013eqa}, also accounted for in the HNV model by the introduction of new contact terms that absorb the unphysical spin-$1/2$ components in the $\Delta$ propagator~\cite{Hernandez:2016yfb}. Extensions of the HNV model to higher energies have been developed by enlarging the resonant content of the model beyond the $\Delta(1232)$~\cite{Hernandez:2013jka,Gonzalez-Jimenez:2016qqq,Kabirnezhad:2017jmf} and by applying Regge phenomenological corrections to the non-resonant contribution~\cite{Gonzalez-Jimenez:2016qqq}. A power counting was introduced in Ref.~\cite{Serot:2012rd} in an effective model with pion, nucleons, $\Delta(1232)$ but also  scalar ($\sigma$) and vector ($\rho$, $\omega$) mesons as degrees of freedom. Next-to-leading order (NLO) (but only tree-level) corrections to weak pion production were investigated. In the dynamical model of Ref.~\cite{Nakamura:2015rta}, the amplitudes are obtained by solving the Lippmann-Schwinger equation in coupled channels, fulfilling Watson's theorem by construction. In this model, PCAC is used to  partially constrain the axial current in terms of the pion-nucleon scattering amplitude fitted to data.  

Chiral perturbation theory (ChPT)~\cite{Weinberg:1978kz,Gasser:1983yg,Gasser:1984gg,Scherer:2012xha}, the effective field theory of QCD at low energies, plays a prominent role in the systematic and model independent study of modern hadronic physics. Initially developed for the description of the interactions among the Goldstone bosons originating from the spontaneous breaking of the $\text{SU(3)}_L\times \text{SU(3)}_R$ chiral symmetry of QCD, it has achieved a remarkable level of precision in the description of a multitude of low-energy observables involving mesons and baryons~\cite{Bijnens:2014lea,Bernard:2007zu}. Amid the large collection of processes successfully described by ChPT we should mention pion photo- and electroproduction off the nucleon. The wealth of precise data available for these reactions has led to an intense theoretical research, reaching a very sophisticated and accurate description of the low energy data; see, e.g., Ref.~\cite{Hilt:2013fda} and references therein for a recent experimental and theoretical review.
 
However, beyond leading-order (LO) tree level amplitudes, the systematic application of ChPT to neutrino-induced pion production has been rare. To our knowledge, it is limited to several low-energy theorems that have been derived for weak pion production, including one-loop corrections, using the heavy-baryon formalism~\cite{Bernard:1993xh}. We report here the first systematic study of weak pion production up to next-to-next-to-leading order (NNLO) in covariant ChPT with nucleons and $\Delta(1232)$. The information gathered in the study in pion production with electromagnetic probes and pion-nucleon scattering within the same framework provides valuable input for  weak pion production. By construction, the amplitudes obtained in ChPT fulfill perturbative unitarity and Watson's theorem. As emphasized in Ref.~\cite{Bernard:1993xh}, ChPT brings about corrections to the axial current that cannot be derived using PCAC. Furthermore, unlike most phenomenological models, it does not require {\it ad hoc} assumptions about the form factors to enforce the (partial) conservation of the (axial) vector current~\cite{Koch:2001ii}. The predictive power of ChPT calculations is limited to the threshold region but nonetheless they can be very valuable for the neutrino cross-section program~\cite{Alvarez-Ruso:2017oui} as a benchmark for phenomenological models that aim to describe weak pion production  in wider energy regions.

This paper is organized as follows. In section~\ref{sec:formalism}, the generic formalism of weak pion production is presented. In section~\ref{sec:HTchpt}, the hadronic tensor is systematically studied in the ChPT framework. Specifically,  we discuss the power counting rule in subsection~\ref{sec:ChPTandPC} and then display all the relevant pieces of the Lagrangian in subsection~\ref{sec:Lag}. The calculation of the 
hadronic transition amplitude and its renormalization are carried out in subsections~\ref{sec:HTA} and \ref{sec:renormalization}, respectively. Section~\ref{sec:results} comprises numerical results: the total cross sections are shown in subsection~\ref{sec:CS} after the parameter values are specified. Pion angular distributions and multipole amplitudes are briefly discussed in subsections~\ref{sec:dis} and~\ref{sec:mul}, respectively. We summarize in section~\ref{sec:summary}. Furthermore, the explicit expressions of the transition amplitude at tree level are compiled in  appendix~\ref{sec:tree.cons}. We also display the axial-vector operators in an alternative basis, well suited for chiral expansions, in appendix~\ref{sec:CESbasis} and he renormalization factors as well as $\beta$ functions are in appendix~\ref{sec:betas}. The amplitudes in the isobaric frame, defined in terms of the Lorentz vector and axial-vector amplitudes and well suited to perform multipole expansions, are shown in appendix~\ref{sec:isobaramp}.

\section{Formalism\label{sec:formalism}}
\subsection{Kinematics, Lorentz and isospin decompositions}

Charged-current weak pion production off the nucleon consists of processes of the type
\bea
\left.
\begin{array}{c}
 \nu_\ell(k_1) \\
 \bar{\nu}_\ell(k_1)   \\
\end{array}
\right\}
+N(p_1)\to
\left.
\begin{array}{c}
\ell^-(k_2) \\
 \ell^+(k_2)   \\
\end{array}
\right\}+N'(p_2)+\pi^b(q)\ ,
\eea
induced either by neutrinos $ \nu_\ell$ or antineutrinos  $\bar{\nu}_\ell$; see Ref.~\cite{Adler:1968tw} for a classic review of electroweak pion production. This reaction is described by the Lorentz-invariant amplitude $\mathcal{T}_{fi}$, which is defined by
\bea
_{\rm out}\langle \ell(k_2)\,N'(p_2)\,\pi^b(q)| \nu_\ell(k_1)\,N(p_1)\rangle_{\rm in}=i(2\pi)^4\delta^{(4)}(k_1+p_1-k_2-p_2-q)\mathcal{T}_{fi} \ .
\eea
In the antineutrino case, one replaces $\nu_\ell\to \bar{\nu}_\ell$ and $\ell^-\to{\ell}^+$ in the above definition. The amplitude $\mathcal{T}_{fi}$ is a function of the following six Mandelstam variables, 
\bea\label{eq:mans}
s&\equiv&(k_1+p_1)^2\ ,\qquad s_1\equiv(k_2+p_2)^2\ ,\qquad s_2\equiv(q+p_2)^2\ ,\nonumber\\
t_1&\equiv&(k_1-k_2)^2\ ,\qquad t_2\equiv(k_1-q)^2\ ,\qquad t\equiv(p_1-p_2)^2\ ,
\eea
which fulfill the constraint
\bea
m_N^2+s+t_1+t_2=t+s_1+s_2\ ,
\eea
where the neutrino mass has been approximated to zero. We work in the isospin limit so the mass of all nucleons (pions) has been set to $m_N$ ($M_\pi$).
Henceforth, $t$ is always given in terms of the other five invariants. 

\begin{figure*}[h!]
\begin{center}
\epsfig{file=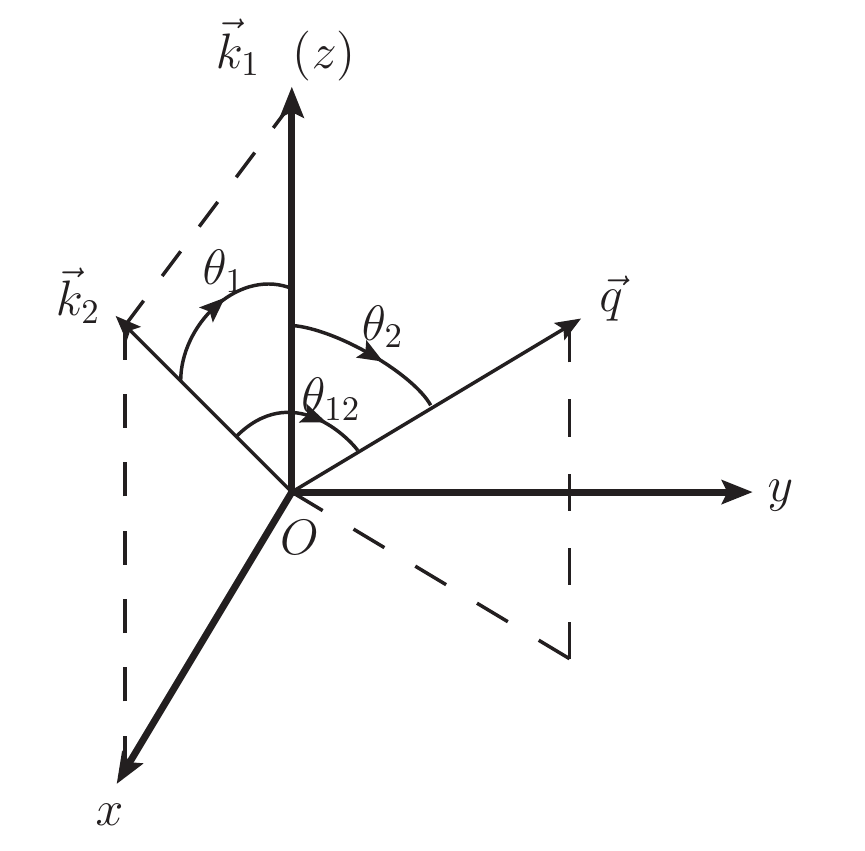,width=0.4\textwidth}
\caption{
Kinematics and reference frame.\label{fig:kinematics}}
\end{center}
\end{figure*}

In the limit $|t_1| \ll M_W^2$, where $M_W$ is the vector W-boson mass, the  scattering amplitude $\mathcal{T}_{fi}$ can be written as
\bea\label{eq:Tfi}
\mathcal{T}_{fi}=\frac{G_F}{\sqrt{2}}|V_{ud}| L^\mu H_\mu\ ,
\eea
where the leptonic and hadronic currents, denoted by $L^\mu$ and $H_\mu$, respectively, are given by
\bea
L^\mu&\equiv&\left\{
\begin{array}{cc}
{\bar{u}_\ell(k_2)\gamma^\mu(1-\gamma_5)u_{\nu_\ell}(k_1)}\ ,& {\rm neutrino }\\
{\bar{v}_{\nu_\ell}(k_1)\gamma^\mu(1-\gamma_5)v_{\ell}(k_2)}\ ,& \text{\rm antineutrino }  \\
\end{array}
\right.
\label{eq:leptonicpart}\\
H_\mu^{ba}&\equiv& \langle N'(p_2)\pi^b(q)|V_{\mu}^a(0)-A_{\mu}^a(0)|N(p_1)\rangle\ ,
\eea
in terms of the isovector vector and axial-vector currents $V_\mu^a$ and $A_\mu^a$; $H_\mu$ depends only on variables $s_2$, $t_1$ and $t$.  Its isospin structure has the form 
\bea
H^{ba}_{\mu}(s_2,t,t_1)&=&\chi_{f}^\dagger\bigg[\frac{1}{2}\{\tau^b,\tau^a\}\,H^+_\mu+\frac{1}{2}[\tau^b,\tau^a]\,H^-_\mu\bigg]\chi_i\ ,
\eea
where $\chi_i$ and $\chi_f$ are isospinors of the initial and final nucleon states, respectively. Furthermore, the Lorentz decomposition reads
\bea\label{eq:lordecom}
H^{\pm}_\mu(s_2,t,t_1)=\sum_{i=1}^8\bar{u}_{N'}(p_2)\bigg\{A_i^{\pm}(s_2,t,t_1)\,\mathcal{O}_{\mu,i}^A+V_i^{\pm}(s_2,t,t_1)\,\mathcal{O}_{\mu,i}^V\bigg\}u_N(p_1)\ ,
\eea
with the Lorentz axial-vector operators \footnote{This simple basis can be easily related to the ones in Ref.~\cite{Adler:1968tw} or Ref.~\cite{Hernandez:2007qq}, if needed.}
\bea\label{eq:OAs}
\mathcal{O}_{\mu,1}^A&=&q_\mu\ ,\qquad \mathcal{O}_{\mu,2}^A=p_{1,\mu}\ ,\qquad \mathcal{O}_{\mu,3}^A=p_{2,\mu} \ ,\nonumber\\
\mathcal{O}_{\mu,4}^A&=&\slashed{q}\,q_\mu\ ,\qquad \mathcal{O}_{\mu,5}^A=\slashed{q}\,p_{1,\mu}\ ,\qquad \mathcal{O}_{\mu,6}^A=\slashed{q}\,p_{2,\mu} \ ,\nonumber\\
\mathcal{O}_{\mu,7}^A&=&\gamma_\mu\,\slashed{q}\ ,\qquad \mathcal{O}_{\mu,8}^A=\gamma_\mu\ ,
\eea
and Lorentz vector operators
\bea\label{eq:OVs}
\mathcal{O}_{\mu,i}^V=\mathcal{O}_{\mu,i}^A\,\gamma_5\ ,\qquad i=1,\cdots,8\ .
\eea
The set of vector operators is complete but they are not independent if the conservation of the vector current is imposed. To be specific, there exist two constraints on $V_i$:
\bea
k\cdot q\, V_1+k\cdot p_1\,V_2+k\cdot p_2\,V_3+(M_\pi^2-2p_1\cdot q)V_7+2m_NV_8&=&0\ ,\nonumber\\
k\cdot q\, V_4+k\cdot p_1\,V_5+k\cdot p_2\,V_6+V_8&=&0\ ,
\eea
with $k\equiv k_1-k_2$. Eventually, once the functions $H^{\pm}_\mu$ are determined, the hadronic transition amplitudes for the various physical weak pion production processes can be readily obtained through
\bea\label{eq:Hphysical}
H_\mu(\nu_\ell\, p\to \ell^-\pi^+p)&=&H_\mu(\bar{\nu}_\ell\, n\to \ell^+\pi^-n)=H^+_\mu-H^-_\mu\ ,\nonumber\\
H_\mu(\nu_\ell\, n\to \ell^-\pi^+n)&=&H_\mu(\bar{\nu}_\ell\, p\to \ell^+\pi^-p)=H^+_\mu+H^-_\mu\ ,\nonumber\\
H_\mu(\nu_\ell\, n\to \ell^-\pi^0p)&=&H_\mu(\bar{\nu}_\ell\, p\to \ell^+\pi^0n)=-\sqrt{2}\,H^-_\mu\ .
\eea

\subsection{Cross section\label{sec:XS}}

Unless  otherwise stated, the energies and momenta are defined in the center-of-mass frame (CM) of the initial (anti)neutrino and nucleon. The directions of pion and lepton three-momenta directions  are specified in the reference frame depicted in Fig.~\ref{fig:kinematics}. By construction, $Oxz$ is the lepton scattering plane.

The total cross section reads
\bea\label{eq:totCS}
\sigma(s)=\frac{1}{(4\pi)^4\sqrt{s}\,|\mathbf{k}_1|}\int_{\omega_\ell^-}^{\omega_\ell^+}{\rm d}\omega_\ell\int_{\omega_\pi^-}^{\omega_\pi^+}{\rm d}\omega_\pi\int_{-1}^{+1}{\rm d}x_1\int_0^\pi {\rm d}\phi_{12}\,|{\cal T}_{fi}|^2,
\eea
where $x_1=\cos\theta_1$ and $\phi_{12}$ is the angle between the $Oxz$ plane and the one spanned by $\vec{k}_2$ and $\vec{q}$. Here, the limits for the lepton energy $\omega_l$ are given by
\bea
\omega_\ell^-=m_\ell\ ,\qquad \omega_\ell^+=\frac{(\sqrt{s}-M_\pi)^2+m_\ell^2-m_N^2}{2(\sqrt{s}-M_\pi)}\ ,
\eea
and the ones for the pion energy $\omega_\pi$ are
\bea
\omega_\pi^{\pm}&=&\frac{1}{2(s-2\omega_\ell\sqrt{s}+m_\ell^2)}\bigg\{(\sqrt{s}-\omega_\ell)(s-2\omega_\ell\sqrt{s}+m_\ell^2+M_\pi^2-m_N^2)\nonumber\\
&&\pm(\omega_\ell^2-m_\ell^2)\sqrt{\left[s-2\omega_\ell\sqrt{s}+m_\ell^2-M_\pi^2-m_N^2\right]^2-4M_\pi^2m_N^2}\bigg\}\ ,
\eea
In the above, $m_\ell$ denotes the outgoing-lepton mass. The invariant amplitude squared can be written as
\bea\label{eq:ccss}
|{\cal T}_{fi}|^2=\frac{G_F^2}{2}|V_{ud}|^2L_{\mu\nu}H^{\mu\nu},\ 
\eea
in terms of the conventional leptonic and hadronic tensors. From Eq.~\eqref{eq:leptonicpart} the leptonic tensor for a neutrino-induced process is given by
\bea
L_{\mu\nu}&=&
{\rm Tr}\big[\slashed{k_1}\gamma_{\mu}(1-\gamma_5)(\slashed{k}_2+m_\ell)\gamma_{\nu}(1-\gamma_5)\big]\nonumber\\
&=&8[k_{1,\mu} k_{2,\nu}+k_{1,\nu} k_{2,\mu}-g_{\mu\nu}k_1\cdot k_2+i\epsilon_{\mu\nu\alpha\beta}k_{1}^{\alpha}k_{2}^{\beta}]\ ,
\eea
with $\epsilon_{0123}=+1$. For the corresponding antineutrino reaction, the term proportional to the fully anti-symmetric tensor gets a minus sign. On the other hand, the hadronic tensor $H_{\mu\nu}$ reads
\bea
H_{\mu\nu}&=&\frac{1}{2}{\rm Tr}\big[(\slashed{p}_1+m_N)\tilde{H}_\mu(\slashed{p}_2+m_N)H_\nu\big]\ ,
\eea
where $
\tilde{H}_\mu=\gamma_0H^\dagger_\mu\gamma_0
$. The hadronic transition amplitudes $H_{\mu}$ are those introduced in Eq.~\eqref{eq:Hphysical}. 

The total cross section is a function of only $s$, so that the other four Mandelstam variables should be expressed in terms of $s$ and the integration variables:
\bea
s_1(s,\omega_\pi)&=&s-2\sqrt{s}\,\omega_\pi+M_\pi^2\ ,\nonumber\\
s_2(s,\omega_\ell)&=&s-2\sqrt{s}\,\omega_\ell+m_\ell^2\ ,\nonumber\\
t_1(s,\omega_\ell,x_1)&=&m_\nu^2+m_\ell^2-2\omega_\nu\,\omega_\ell+2|\vec{k}_1||\vec{k}_2|\,x_1\ ,\nonumber\\
 t_2(s,\omega_\ell,\omega_\pi,x_1,\phi_{12})&=&m_\nu^2+M_\pi^2-2\omega_\nu\,\omega_\pi+2|\vec{k}_1||\vec{q}|\,x_2\ ,
\eea
where {$x_i\equiv \cos\theta_i$ and} the moduli of the three momenta are 
\bea
|\vec{q}|&=&\sqrt{\omega_\pi^2-M_\pi^2}\ ,\qquad |\vec{k}_2|=\sqrt{\omega_l^2-m_l^2}\ ,\qquad|\vec{k}_1|=\omega_\nu \ ,
\eea
with $\omega_\nu={(s-m_N^2)}/{(2\sqrt{s})}$. Furthermore, $x_2=x_1x_{12}+\sqrt{(1-x_1^2)(1-x_{12}^2)}\cos\phi_{12}$, and $x_{12}$ is obtained from
\bea
|\vec{k}_2||\vec{q}|\,x_{12}=\frac{1}{2}(m_\ell^2+M_\pi^2-m_N^2+s)-\sqrt{s}(\omega_\ell+\omega_\pi)+\omega_\ell\,\omega_\pi\ .
\eea
The invariant $s$ can be related to the energy of the neutrino in the laboratory frame, $E_\nu$, by
\bea
s=m_N^2+2m_N\,E_\nu\ ,
\eea
so that the total cross section can expressed as a function of $E_\nu$.

\section{Systematic analysis of the hadronic tensor in ChPT\label{sec:HTchpt}}

In this section, the different ingredients required to obtain the hadronic current in ChPT are presented. 

\subsection{Power counting\label{sec:ChPTandPC}}

As an expansion in powers of momenta and light-quark masses, ChPT relies on a hierarchy of the contributions (diagrams) known as power counting. The presence of matter fields as explicit degrees of freedom introduces new scales that do not vanish in the chiral limit, causing the presence of power counting breaking (PCB) terms~\cite{Gasser:1987rb}  in the diagrams with loops. To remedy this problem, various approaches have been proposed in the past thirty years: e.g., the heavy baryon (HB) formalism~\cite{Jenkins:1990jv,Bernard:1992qa}, the infrared regularization (IR) prescription~\cite{Ellis:1997kc,Becher:1999he}, and the extended-on-mass-shell (EOMS) scheme~\cite{Gegelia:1999gf,Gegelia:1999qt,Fuchs:2003qc}.\footnote{See also Refs.~\cite{Bernard:2007zu,Geng:2013xn} for further discussion on this topic.}

For ChPT in the one-baryon sector, denoted in short as BChPT, the EOMS scheme has proven to be a very effective tool. It is covariant and preserves the analytic structure of the calculated physical quantities with correct power counting. When the proper limits are taken, EOMS reproduces the results obtained using the HB  or the IR formalisms but usually offers a faster chiral convergence because covariance and the analytic structure of the loops are maintained~\cite{Geng:2010df,Geng:2008mf,MartinCamalich:2010fp}. Due to the above-mentioned facts, the EOMS scheme is  gaining a widespread acceptance and has been applied to many relevant processes, e.g. pion-nucleon scattering~\cite{Alarcon:2012kn,Chen:2012nx,Yao:2016vbz,Siemens:2016hdi} and pion photoproduction~\cite{Hilt:2013uf,Blin:2014rpa,Blin:2016itn}, among others. It  has also been used to describe heavy-light systems~\cite{Geng:2010vw,Yao:2015qia,Du:2017ttu}. Furthermore, there have been attempts to create a new framework based on EOMS to extend the applicability beyond the low-energy region but restricted to small scattering angles~\cite{Epelbaum:2015vea}.

The explicit inclusion in BChPT  of  baryon states heavier than the nucleon, such as the $\Delta$ resonance, is not trivial. The $\Delta(1232)$ excitation is the lightest baryon resonance, located only $\sim 200$~MeV above the $\pi N$ threshold, and hence crucial for a good description of the $\pi N$ physics even at low energies. In BChPT with $\Delta(1232)$, apart from the external momenta $p$ and the pion mass $M_\pi$, an additional small parameter appears, namely the mass difference $\delta=m_\Delta-m_N\sim 300$~MeV. Different assumptions about the expansion parameters lead to different power-counting rules. In the small scale expansion (SSE) scheme proposed in Refs.~\cite{Hemmert:1996xg,Hemmert:1997ye}, both $\delta$ and $M_\pi$ are counted as $\mathcal{O}(p)$. Instead, in the so-called $\delta$-counting, developed in Ref.~\cite{Pascalutsa:2002pi}, a different counting, $\delta\sim \mathcal{O}(p^{\frac{1}{2}})$, is introduced in order to preserve the hierarchy $p/\Lambda_{\chi {\rm SB}} \sim M_\pi/\Lambda_{\chi {\rm SB}} \sim (\delta/\Lambda_{\chi {\rm SB}})^2$, with $\Lambda_{\chi {\rm SB}}\sim 1$~GeV being the chiral symmetry breaking scale.

In the present work, we are interested in the energy range from the production threshold $E_\nu^{\rm thr.}$ ($\simeq 276.5$~MeV for $\ell=\mu$)  to $E_\nu^{\rm max}\sim E_\nu^{\rm thr.} +M_\pi$ ($\simeq 415$~MeV for $\ell=\mu$). With such a choice, $Q^2 \equiv -t_1$ is always smaller than $0.02$~GeV$^2$ and the pion momentum is smaller than $0.18$~GeV. Furthermore, the invariant mass of the final hadronic $\pi N$ system, denoted as $W \equiv \sqrt{s_2}$, is $\leq1.18$~GeV, well below the $\Delta$-resonance peak.
Hence, we prefer to employ the $\delta$-counting rule. Specifically, for a given Feynman diagram with $L$ loops, $V^{(k)}$ vertices of $\mathcal{O}(p^k)$, $I_\pi$ internal pions, $I_N$ nucleon propagators and $I_\Delta$ $\Delta$-propagators, its chiral dimension $D$ is obtained according to the rule
\bea\label{eq:rule}
D=4L+\sum_k kV^{(k)}-2I_\pi-I_N-\frac{1}{2}I_\Delta.
\eea
Here, we aim to perform a calculation of the hadronic transition amplitude up to  the chiral order $\mathcal{O}(p^3)$, i.e. $\mathcal{O}({p^{D}}/{\Lambda^D_{\chi {\rm SB}}})$ with $D=3$.

\subsection{Chiral effective Lagrangians\label{sec:Lag}}
Given our working accuracy and according to the power counting rule~\eqref{eq:rule}, the following chiral Lagrangians are needed for our calculation,
\bea
\mathcal{L}_{\rm eff}=\sum_{i=1}^2\mathcal{L}_{\pi\pi}^{(2i)}+\sum_{j=1}^3\mathcal{L}_{\pi N}^{(j)}+\sum_{k=1}^2\left[{\cal L}^{(k)}_{\pi\Delta}+{\cal L}^{(k)}_{\pi N\Delta}\right]\ ,
\eea
where superscripts represent chiral orders while subscripts denote the relevant degrees of freedom. For clarity, the effective Lagrangian is classified in three parts: the purely pionic sector, the pion-nucleon sector and the one involving $\Delta$ resonances.

\subsubsection{Pionic interactions}
The required terms in the  purely pionic sector are given by~\cite{Gasser:1983yg,Gasser:1987rb}
\bea
\mathcal{L}_{\pi\pi}^{(2)}&=&\frac{F^2}{4}{\rm Tr}[D_\mu U {(D^\mu U)}^\dagger+\chi U^\dagger+U\chi^\dagger]\ ,\label{eq:LOLagpipi}\\
\mathcal{L}_{\pi\pi}^{(4)}&=&\frac{\ell_3+\ell_4}{16}[{\rm Tr}(\chi U^\dagger+U\chi^\dagger)]^2+\frac{\ell_4}{8}{\rm Tr}[D_\mu U(D^\mu U)^\dagger]{\rm Tr}[\chi U^\dagger+U\chi^\dagger]\nonumber\\
&&+i\frac{\ell_6}{2}{\rm Tr}[F^L_{\mu\nu}(D^\mu U)^\dagger D^\nu U]\ ,\label{eq:LagPi4}
\eea
where $F^L_{\mu\nu}=\partial_\mu l_\nu-\partial_\nu l_\mu-i[l_\mu,l_\nu]$ is the left-handed field-strength tensor; $l_\mu=-{g_W}\, V_{ud} {l_\mu^a\tau^a}/{2}$ is the left-handed external field and $\tau^a$ $(a=1,2,3)$ are the Pauli matrices.\footnote{We identify $l_\mu^1 =\mathcal{W}_\mu^1$, $l_\mu^2=\mathcal{W}_\mu^2$ and $l_\mu^3=0$, to which the physical weak-boson fields $\mathcal{W}_\mu^\pm$ are related via $\mathcal{W}_\mu^\pm=(\mathcal{W}^1_\mu\mp i\,\mathcal{W}_\mu^2)/\sqrt{2}$. Note that, to be consistent with Eq.~\eqref{eq:Tfi}, we always factorize out the combination $-{g_W}/(2 \sqrt{2})\, V_{ud}$ from the hadronic transition amplitude $H_\mu$ calculated in subsection~\ref{sec:HTA} . Furthermore, the factor $g_W/(2 \sqrt{2})$, together with an identical one from the lepton sector, is absorbed in the Fermi constant as $G_F=\sqrt{2}\frac{{g^2_W}}{8 M_W^2}$, where $M_W$ denotes the mass of the vector $W$ boson. } 
Here $\chi={\rm diag}\{M^2,M^2\}$ is the mass matrix with $M$ being the pion mass in the isospin limit. ${\rm Tr[\cdots]}$ denotes the trace in flavor space. Furthermore, $F$ is the pion decay constant in the chiral limit and $\ell_{3,4,6}$ are mesonic low-energy constants (LECs). The Goldstone pion fields are collected in the $2\times2$ matrix $U$ 
\bea
U=u^2={\rm exp}\bigg(\frac{i\tau^b\pi^b}{F}\bigg)\ ,
\quad D_\mu U=\partial_\mu U+i U\,l_\mu\ ,
\eea
where the corresponding covariant derivative has also been defined.

\subsubsection{Interactions with nucleons}
The relevant terms describing the interactions between pions, or external fields $l_\mu$, and nucleons read~\cite{Fettes:2000gb}
\bea
\mathcal{L}_{\pi N}^{(1)}&=&\bar{\Psi}_{N}\{i\slashed{D}-m+\frac{g}{2}\slashed{u}\gamma_5\}\Psi_N\ ,\label{eq:LOLagpiN}\\
\mathcal{L}_{\pi N}^{(2)}&=&\bar{\Psi}_{N}\bigg\{c_1{\rm Tr}[\chi_+]-\frac{c_2}{4m^2}{\rm Tr}[u^\mu u^\nu](D_\mu D_\nu+h.c.)+\frac{c_3}{2}{\rm Tr}[u^\mu u_\mu]\nonumber\\
&+&\bigg[\frac{i\,c_4}{4}[u_\mu,u_\nu]+\frac{c_6}{8m}F_{\mu\nu}^+\bigg]\sigma^{\mu\nu}\bigg\}\Psi_{N}\ ,\label{eq:LagN2}\\
{\cal L}_{\pi N}^{(3)}&=&\bar{\Psi}_N\bigg\{-\frac{d_1}{2m}\big([u_\mu,[D_\nu,u^\mu]]D^\nu+h.c.)-\frac{d_2}{2m}\big([u_\mu,[D^\mu,u_\nu]]D^\nu+h.c.)\nonumber\\
&+&\frac{d_3}{12m^3}([u_\mu,[D_\nu,u_\lambda]](D^\mu D^\nu
D^\lambda+sym.)+h.c.\big)+\frac{d_5}{2m}(i[\chi_-,u_\mu]D^\mu+h.c.)\nonumber\\
&+&\frac{d_6}{2m}(i[D^\mu,\tilde{F}_{\mu\nu}^+]D^\nu+h.c.)+\frac{d_8}{2m}(i\epsilon^{\mu\nu\alpha\beta}{\rm Tr}[\tilde{F}_{\mu\nu}^+u_\alpha] D_\beta+h.c.)\nonumber\\
&+&\frac{d_{14}}{4m}(i\sigma^{\mu\nu}{\rm Tr}[[D_\lambda,u_\mu]u_\nu]
D^\lambda+h.c.)+\frac{d_{15}}{4m}(i\sigma^{\mu\nu}{\rm Tr}[ u_\mu[D_\nu,u_\lambda]]
D^\lambda+h.c.)\nonumber\\
&+&\frac{d_{16}}{2}\gamma^\mu\gamma^5{\rm Tr}[\chi_+]
u_\mu+\frac{d_{18}}{2}i\gamma^\mu\gamma^5[D_\mu,\chi_-]-\frac{d_{20}}{8m^2}(i\gamma^\mu\gamma_5[\tilde{F}_{\mu\nu}^+,u_\lambda]D^{\lambda\nu}+h.c.)\nonumber\\
&+&\frac{d_{21}}{2}i\gamma^\mu\gamma_5[\tilde{F}_{\mu\nu}^+,u^\nu]+\frac{d_{22}}{2}\gamma^\mu\gamma_5[D^\nu,F^-_{\mu\nu}]+\frac{d_{23}}{2}\gamma_\mu\gamma_5\epsilon^{\mu\nu\alpha\beta}{\rm Tr}[ u_\nu F_{\alpha\beta}^-]\bigg\}\Psi_N\ ,\label{eq:LagN3}
\eea
with the nucleon doublet $\Psi_N=(p,n)^T$. Here, $m$ and $g$ are the nucleon mass and axial charge in the chiral limit. The LECs $c_i$ and $d_j$ have units of GeV$^{-1}$ and GeV$^{-2}$, respectively. The involved chiral blocks are given by
\bea
u_\mu&=& iu^\dagger\partial_\mu U u^\dagger+i\, u\, l_\mu u^\dagger\ ,\quad
\Gamma_\mu= \frac{1}{2}[u^\dagger,\partial_\mu u]-\frac{i}{2}\, u\, l_\mu u^\dagger\ ,\quad  D_\mu=\partial_\mu+\Gamma_\mu\ ,\\
\chi_\pm&=&u^\dagger\chi u^\dagger\pm u\chi^\dagger u\ ,\quad
F_{\mu\nu}^\pm=\pm u F^L_{\mu\nu}u^\dagger\ , \quad \tilde{F}_{\mu\nu}^+={F}_{\mu\nu}^+-\frac{1}{2}{\rm Tr}[{F}_{\mu\nu}^+]\ .
\eea
In practice, the Levi-Civita tensor can be expressed in terms of Dirac gamma matrices:  
$
\epsilon^{\mu\nu\alpha\beta}=-\frac{i}{8}[\{[\gamma^\mu,\gamma^\nu],\gamma^\alpha\},\gamma^\beta]\gamma_5
$. 
In such a manner, the Lorentz structure of the hadronic transition amplitude can be readily  expressed in terms of the operators given in Eqs.~\eqref{eq:OAs} and~\eqref{eq:OVs}.

\subsubsection{Interactions with \texorpdfstring{$\Delta$}{Delta}}
The $\Delta$-resonance is a state of spin-$3/2$, which can be represented by a vector-spinor ${\Psi}^{\mu}$ in the Rarita-Schwinger formalism~\cite{Rarita:1941mf}. It is also a field of isospin-$3/2$,  thus, it can be described by a vector-spinor isovector-isospinor field ${\Psi}^{i,\mu}$, with $\mu$ and $i$ being the Lorentz vector and isovector indices, respectively. 
We refer the reader to Ref.~\cite{Hemmert:1997ye} for the so-called isospurion formulation where the relations between the field ${\Psi}^{i,\mu}$ and the physical $\Delta$(1232) states, $\Delta^{++}$, $\Delta^{+}$, $\Delta^{0}$ and $\Delta^{-}$, are presented.
The interactions of $\Delta$ resonances with pions read
\bea
{\cal L}^{(1)}_{\pi\Delta}&=&\bar{\Psi}^{i,\mu}\xi^{\frac{3}{2}}_{ij}
\left(i \gamma_{\mu\nu\alpha}{D}^{\alpha,jk}-m_\Delta\gamma_{\mu\nu}\delta^{jk}\right)
\xi^{\frac{3}{2}}_{kl}{\Psi}^{l,\nu}\  ,
 \label{Dlagr0}
 \\
{\cal L}^{(2)}_{\pi\Delta}&=& \bar{\Psi}^{i,\mu}\xi^{\frac{3}{2}}_{ij}
\left(a_1{\rm Tr}[\chi_+]\delta^{jk}
g_{\mu\nu}\right)\xi_{kl}^{\frac{3}{2}}\Psi^{l,\nu}\ ,
\label{DLagr}
\eea
where $m_\Delta$ is the $\Delta$ bare mass and $a_1$ a bare coupling constant; the covariant derivative is defined by
\bea
{\cal D}_{\mu,ij}&=&(\partial_\mu+\Gamma_\mu)\delta_{ij}-i\epsilon_{ijk}{\rm Tr}[\tau^k\Gamma_{\mu}]\ .
\eea
Furthermore, $\xi^\frac{3}{2}_{ij}=\delta_{ij}-\frac{1}{3}\tau_i\tau_j$ is the isospin-$3/2$ projection operator; the Dirac matrices with multiple Lorentz indices are defined as
\bea
\gamma_{\mu\nu\alpha}&=&\frac{1}{4}\{[\gamma_\mu,\gamma_\nu],\gamma_\alpha\}\ ,\qquad \gamma_{\mu\nu}=\frac{1}{2}[\gamma_\mu,\gamma_\nu]\ .
\eea
Finally, the effective Lagrangian for pion-nucleon-$\Delta$ interaction has the form~\cite{Hemmert:1996xg,Hemmert:1997ye,Fettes:2000bb}
\bea
{\cal L}^{(1)}_{\pi N\Delta} &=&h_A\bar{\Psi}^{i,\alpha}\xi_{ij}^{\frac{3}{2}}\, \omega_\alpha^j\Psi_N+h.c.\,,\label{eq:LOLagpiND}\\
{\cal L}^{(2)}_{\pi N\Delta} &=& \bar{\Psi}^{i,\alpha}\xi_{ij}^{\frac{3}{2}}\bigg\{-i\frac{\,b_1}{2}F_{\alpha\beta}^{+,j}\gamma_5\gamma^\beta+ib_2\,F_{\alpha\beta}^{-,j}\gamma^\beta+i\,b_3\,\omega_{\alpha\beta}^j\gamma^\beta\nonumber\\
&+&i\,\frac{b_7}{m}F_{\alpha\beta}^{-,j}\,i D^\beta+i\frac{b_8}{m}\omega_{\alpha\beta}^j\,i D^\beta\bigg\}\Psi_N+h.c.\ ,
\eea
where $h_A$ denotes the LO axial coupling constant, $b_k$ are NLO LECs, and the chiral blocks with isovector index $i$ are defined as
\bea
F_{\mu\nu}^{\pm,i}=\frac{1}{2}{\rm Tr}[\tau^i F_{\mu\nu}^\pm]\ ,\quad \omega_\mu^i=\frac{1}{2}{\rm Tr}[\tau^i u_\mu]\ ,\quad \omega_{\mu\nu}^i=\frac{1}{2}{\rm Tr}[\tau^i [D_\mu,u_\nu]]\ .
\eea
In fact,  as pointed out in Ref.~\cite{Jiang:2017yda}, the $b_2$ and $b_7$ terms can be eliminated thanks to the identity, $F^-_{\mu\nu}=[D_\mu, u_\nu]-[D_\nu, u_\mu]$. Furthermore, the $b_3$ and $b_8$ terms are redundant too~\cite{Jiang:2017yda, Long:2010kt}, which has been explicitly checked in $\pi N$ scattering~\cite{Yao:2016vbz}, 
showing that their contributions can be absorbed in the LO $\Delta$-exchange and contact terms. Therefore, for ${\cal L}^{(2)}_{\pi N\Delta}$,  we only need to take the $b_1$ term into consideration.

\subsection{Hadronic transition amplitudes\label{sec:HTA}}

The tree-level diagrams relevant to our calculation up to $\mathcal{O}(p^3)$ are depicted in Fig.~\ref{fig:treetopo}. They are labeled according to the scheme shown in Table~\ref{tab:tree} in Appendix~\ref{sec:tree.cons}. Therein, the chiral order of each tree-level contribution is specified, as well, for convenience. The explicit expressions for the corresponding amplitudes are listed diagram by diagram in this appendix.

\begin{figure*}[ht]
\begin{center}
\epsfig{file=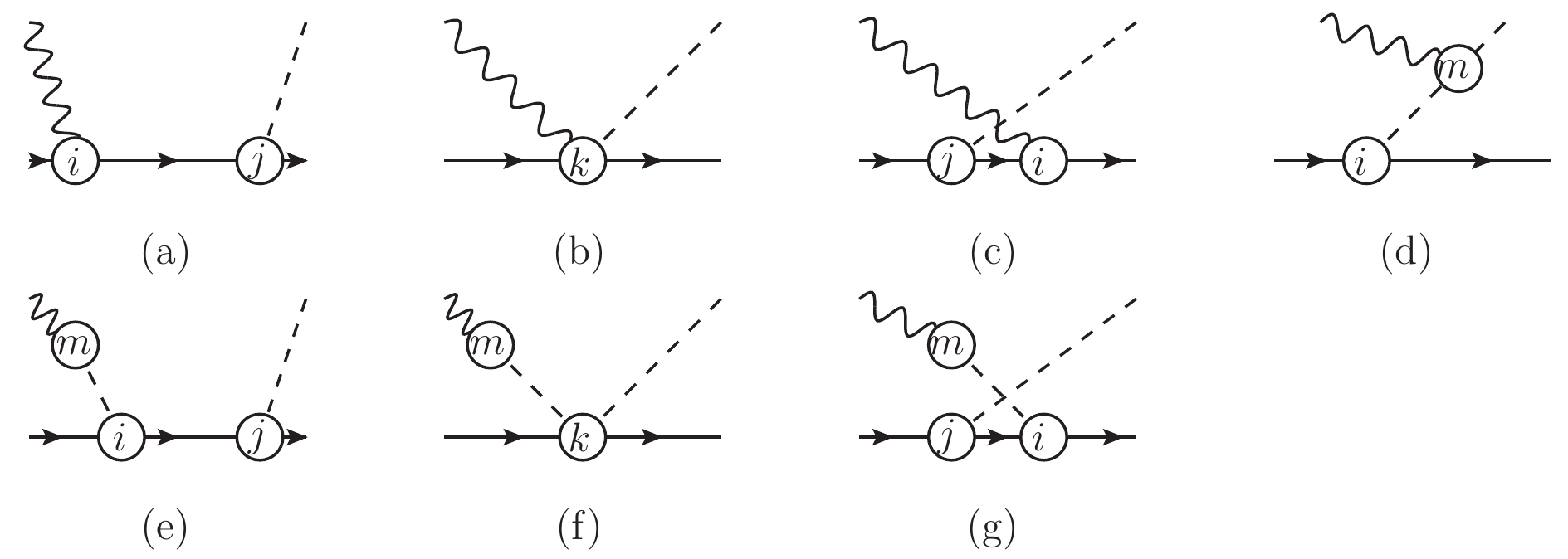,width=0.78\textwidth}
\caption{Topologies of tree-level diagrams. The solid, dashed and wiggled lines represent nucleons, pions and left-hand currents. The letters in the circles mark the possible chiral orders of the vertices. Diagrams with $\Delta$-exchange are obtained by replacing internal nucleon lines by $\Delta$ propagators. Diagrams with mass insertions in the internal pion, nucleon and $\Delta$ propagators are not shown explicitly. \label{fig:treetopo}}
\end{center}
\end{figure*}

In Fig.~\ref{fig:treetopo}, the diagrams with mass insertions in the internal pion, nucleon and $\Delta$ propagators are not shown. Such amplitudes with mass insertions in internal nucleon and $\Delta$ lines, which are generated by terms proportional to the $c_1$ term in $\mathcal{L}_{\pi N}^{(2)}$ and the $a_1$ term in $\mathcal{L}_{\pi \Delta}^{(2)}$, can be taken into account  by the following replacement in the nucleon and $\Delta$ propagators:
\bea
m&\to& m_2 = m-4c_1 M^2\ ,\nonumber\\
m_\Delta&\to& m_{\Delta,2}=m_{\Delta}-4a_1 M^2\ .
\eea
On the other hand, the insertions in pion propagators, generated by the $l_3$ and $l_4$ terms in $\mathcal{L}_{\pi \pi}^{(4)}$, contain momentum-dependent pieces. Hence, their contribution can not be incorporated as in the nucleon and $\Delta$ cases. Instead, the contribution of a diagram with one insertion in a pion line results from the substitution 
\bea
H_\mu^{\pm} \longrightarrow \xi(q_\pi^2) H_\mu^{\pm}\ ,
\eea
with
\bea
\xi(q_\pi^2) = -\frac{2M^2}{F^2}\left(l_4+l_3\frac{M^2}{M^2-q_\pi^2}\right),
\eea
where $q_\pi$ is the momentum transferred in the pion propagator. Note that, up to the order we are working in, the pion-insertions for diagrams $T^D_{12}$, $T^E_{112}$, $T^F_{12}$ and $T^G_{112}$ need to be taken into consideration only once, since $\xi(q_\pi^2)$ is of order $\mathcal{O}(p^2)$.

\begin{figure*}[ht]
\begin{center}
\epsfig{file=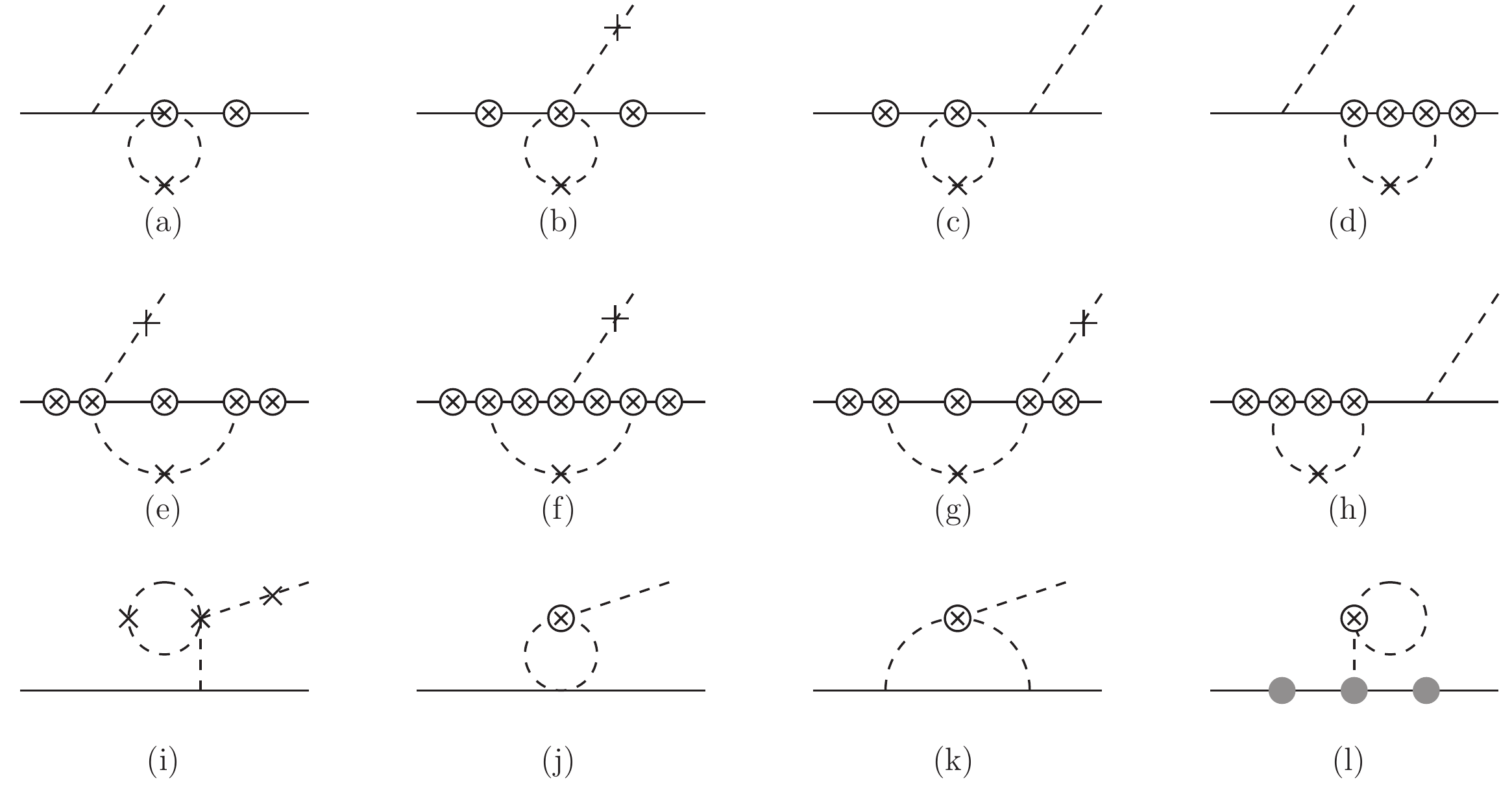,width=0.9\textwidth}
\caption{Topologies from which one-loop diagrams are generated.
Topologies leading to corrections on the external pion and nucleon legs are not shown because the corresponding contributions are taken into account by wave-function renormalization. The solid lines represent nucleons, while the dashed ones stand for the pions. Vertices with crosses, circles and grey dots denote positions at which incoming left-hand currents, incoming pions and outgoing pions, respectively, can be inserted. Incoming pions are always coupled to    
left-hand currents.
\label{fig:looptopo}}
\end{center}
\end{figure*}

\begin{figure*}[ht]
\begin{center}
\epsfig{file=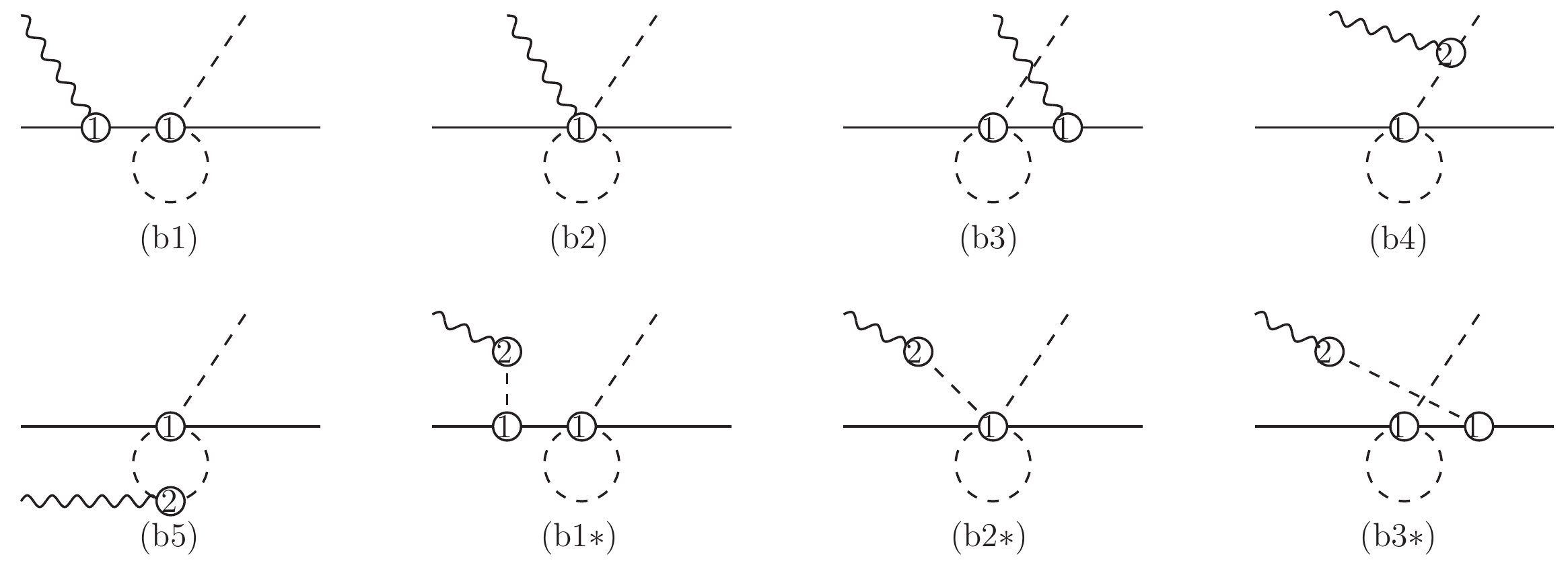,width=0.85\textwidth}
\caption{One-loop diagrams generated from topology (b) of Figure~\ref{fig:looptopo}. The solid, dashed and wiggled lines represent nucleons, pions and left-hand currents. Circled numbers mark the chiral orders of the vertices. \label{fig:out2cc}}
\label{fig:example.loop}
\end{center}
\end{figure*}

For the calculation of loop contributions, we need all the diagrams generated from
the topologies shown in Fig.~\ref{fig:looptopo}. In total, there are 89 diagrams. An example of how to generate them from topology (b) of Fig.\ref{fig:looptopo} is shown in Fig.~\ref{fig:example.loop}. The calculation of these one-loop amplitudes is straightforward but yields lengthy analytical expressions, which we do not show explicitly here\footnote{The simpler expressions of the one-loop contributions obtained for pion photoproduction can be found in Ref~\cite{Blin:2014rpa}}, but can be obtained from the authors upon request.
Finally, the contributions of diagrams corresponding to loop corrections on the external legs are included through wave function renormalization, which is discussed in the next section.

\subsection{Renormalization\label{sec:renormalization}}

In the above subsection, we have described the calculation of the hadronic transition amplitudes up to $\mathcal{O}(p^3)$, corresponding to the Feynman diagrams excluding corrections at external pion and nucleon legs.  In fact, the sum of all their contributions yields the amputated amplitude, $\hat{H}_\mu$, for which the superscripts `$\pm$' are suppressed for brevity. According to the Lehmann-Symanzik-Zimmermann (LSZ) reduction formula~\cite{Lehmann:1954rq}, the full amplitude is related to the amputated one through
\bea\label{eq:wfr}
H_\mu(s_2,t,t_1) = {\cal Z}_\pi^{\frac{1}{2}} \mathcal{Z}_N \hat{H}_\mu(s_2,t,t_1)\ ,
\eea
where $\mathcal{Z}_\pi$ and $\mathcal{Z}_N$ are wave function renormalization functions of the pion and nucleon fields, respectively. Their explicit expressions are given in Appendix~\ref{sec:betas}. 

In the full amplitude, the loop contributions are evaluated using dimensional regularization. The ultraviolet (UV) divergences stemming from the loops are subtracted using the modified minimal subtraction ($\overline{\rm MS}$-1) scheme and absorbed by the LECs appearing in the counter terms generated by the effective Lagrangian. That is, we split the bare LECs in the following way,
\bea\label{eq:UVshift}
X = X^r +\frac{\beta_X}{16\pi^2} R\ , \quad  X\in\{m,g, c_i, d_j, l_k\}\ ,
\eea
where $R={2}/{(d-4)}+\gamma_E-1-\ln(4\pi)$, $d$  the number of space-time dimension, and $\gamma_E$  the Euler constant.  We refer to the effective Lagrangians, in Eqs.~\eqref{eq:LagN2}, \eqref{eq:LagN3} and \eqref{eq:LagPi4}, for the values of the indices $i, j, k$. Furthermore, $\beta_X$ are beta functions.

As already mentioned in the beginning of this section, there exist PCB terms due to the appearance of nucleon internal lines in the loop diagrams. To restore the power counting, we apply the EOMS scheme.  Therefore, after the cancellation of the UV divergences, one has to perform additional finite shifts for the $\mathcal{O}(p)$ and $\mathcal{O}(p^2)$ UV renormalized LECs as
\bea\label{eq:finiteshift}
X^r = \tilde{X} +\frac{m\,\tilde{\beta}_X}{16\pi^2F^2} \ , \quad  X\in\{m,g, c_i\}\ ,
\eea
with $\tilde{\beta}_X$ being the beta functions for this finite renormalization.

The verification of the cancellation of UV divergences and PCB terms is delicate. The vector and axial-vector operators given in Eqs~\eqref{eq:OAs} and~\eqref{eq:OVs} are not well suited to perform a chiral expansion, due to the fact that sometimes the chiral order of their combination is underestimated. For instance, the chiral orders of $\MOA_{\mu,5}$ and $\MOA_{\mu,6}$ are both assigned to be $\mathcal{O}(p)$. Consequently, the combination $\MOA_{\mu,5}-\MOA_{\mu,6}$ is naively counted as $\mathcal{O}(p)$. However, its actual chiral order should be $\mathcal{O}(p^2)$, since $(p_{1,\mu}-p_{2,\mu})$ gives an additional contribution of $\mathcal{O}(p)$. Therefore, to overcome such issues during renormalization, we have chosen a chiral-expansion-suited (CES) basis, see Eq.~\eqref{eq:CESbasisOA} and Eq.~\eqref{eq:CESbasisOV}  in Appendix~\ref{sec:CESbasis}. Another advantage of the CES basis is that vector current conservation is automatically implemented. With the help of the CES basis, we remove the UV divergences and PCB terms order by order in the chiral expansion, and obtain the explicit expressions for the $\beta$ functions, namely $\beta_X$ and $\tilde{\beta}_X$ in Eqs~\eqref{eq:UVshift} and~\eqref{eq:finiteshift}, which are relegated to Appendix~\ref{sec:betas}. 

All the parameters in the renormalized full amplitude are UV finite. For practical convenience, we write $F$, $M$, $\tilde{m}$ and $\tilde{g}$ in terms of their corresponding physical values, $F_\pi$, $M_\pi$, $m_N$ and $g_A$ by using the relations specified in Eq.~\eqref{eq:renfactors1} and Eq.~\eqref{eq:renfactors2}. 
The terms of $\mathcal{O}(p^4)$ and higher orders  generated by the above substitutions, as well as by the wave function renormalization in Eq.~\eqref{eq:wfr} are neglected.

\section{Numerical results and discussion\label{sec:results}}
\subsection{Low energy constants\label{sec:LECs}}

The available data for neutrino-induced charged-current single pion production on nucleons at low energies are very scarce. In fact, they are limited to the early experimental measurements at the ANL~\cite{Radecky:1981fn,Barish:1978pj} and the BNL~\cite{Kitagaki:1986ct,Kitagaki:1990vs} hydrogen- and deuterium-filled bubble chambers. These data have been recently reanalyzed for the flux uncertainty in Ref.~\cite{Rodrigues:2016xjj}. Muon neutrino beams were used for both ANL and BNL  with average energies around $1$~GeV and $1.6$~GeV, respectively. Although events for all allowed channels induced by muon neutrinos were detected, almost all the data are beyond the energy region where ChPT is  expected to be valid. This is also the case for the data on muon antineutrino-induced processes measured at CERN-PS~\cite{Bolognese:1979gf}. Therefore, the task of fixing the unknown LECs present in 
the hadronic transition amplitudes calculated above by fitting the above mentioned $\nu-\bar{\nu}$ data is unattainable. Nonetheless, most of the required LECs are known, as they have been obtained in the analysis of other processes or physical quantities. We take their values from the studies of $\pi N$ scattering~\cite{Alarcon:2012kn,Chen:2012nx,Yao:2016vbz}\footnote{A recent determination of some of the LECs has been performed in Ref.~\cite{Siemens:2016jwj} by making use of $\pi N$ threshold and subthreshold parameters, instead of partial wave phase shifts.} and the axial radius of the nucleon~\cite{Yao:2017fym}, which used the EOMS scheme as in the present calculation.

For the parameters appearing in the LO Lagrangians, i.e., $\mathcal{L}_{\pi\pi}^{(2)}$, $\mathcal{L}_{\pi N}^{(1)}$, $\mathcal{L}_{\pi\Delta}^{(1)}$ and $\mathcal{L}_{\pi N\Delta}^{(1)}$ [Eqs.~\eqref{eq:LOLagpipi}, \eqref{eq:LOLagpiN}, \eqref{Dlagr0} and \eqref{eq:LOLagpiND}], the values of their corresponding physical counterparts are set to~\cite{Patrignani:2016xqp,Bernard:2012hb}
\bea\label{eq:phyval}
&&F_\pi = 92.21~\text{MeV}\ , \quad g_A=1.27\ ,  \quad h_A=1.43\pm0.02\ ,\no
&&M_\pi =138.04~\text{MeV}\ ,\quad m_N=938.9~\text{MeV} \ ,\quad m_\Delta=1232~\text{MeV}\ ,
\eea
where $h_A$ is determined from the strong decay width of $\Delta\to\pi N$ ($\Gamma_\Delta^{\rm str.}=118\pm 2$~MeV~\cite{Patrignani:2016xqp}). See Ref.~\cite{Bernard:2012hb} for details.

\begin{table}[h]
\caption{Values of the LECs determined from other processes. Details on the different sources are explained in the text. Here $d^r_{1+2}=d^r_1+d^r_2$ and $d^r_{14-15}=d^r_{14}-d^r_{15}$.}\label{tab:LECs}
\vspace{-0.5cm}
\bea
\begin{array}{cc|rc}
\hline\hline
&{\rm LEC}& \text{Value}&\text{Source}\\
\hline
\mathcal{L}_{\pi\pi}^{(4)}&\bar{\ell}_6&16.5\pm1.1&\text{$\langle r^2\rangle_\pi$~\cite{Gasser:1983yg}}\\
\hline
\multirow{5}{*}{\text{$\mathcal{L}_{\pi N}^{(2)}$}}&\tilde{c}_1&-1.00\pm 0.04&\multirow{4}{*}{\text{$\pi N$~scattering~\cite{Alarcon:2012kn}}}\\
&\tilde{c}_2&1.01\pm0.04&\\
&\tilde{c}_3&-3.04\pm0.02&\\
&\tilde{c}_4&2.02\pm0.01&\\
\cline{4-4}
&\tilde{c}_6&1.35\pm0.04&\text{$\mu_p$ and $\mu_n$~\cite{Bauer:2012pv,Patrignani:2016xqp}}\\
\hline
\multirow{6}{*}{\text{$\mathcal{L}_{\pi N}^{(3)}$}}&d_{1+2}^r&0.15\pm 0.20&\multirow{5}{*}{\text{$\pi N$~scattering~\cite{Alarcon:2012kn}}}\\
&d_3^r&-0.23\pm0.27&\\
&d_5^r&0.47\pm0.07&\\
&d_{14-15}^r&-0.50\pm0.50&\\
&d_{18}^r&-0.20\pm0.80&\\
\cline{4-4}
&d_{22}^r&0.96\pm0.03&\text{$\langle r_A^2\rangle_N$~\cite{Yao:2017fym}}\\
\hline
\mathcal{L}_{\pi N\Delta}^{(2)}&b_1&(4.98\pm0.27)/m_N&\text{$\Gamma_\Delta^{\rm em}$~\cite{Bernard:2012hb}}\\
\hline\hline
\end{array}\nonumber
\eea
\end{table}

In the higher-order effective Lagrangians relevant to our calculation, there are in total 22 LECs. Three of them, $\ell_3^r$, $\ell_4^r$ and $d_{16}^r$, become irrelevant after the procedure of renormalization and replacement of the LO parameters by their physical ones as discussed in the previous section. Furthermore, as shown in Table~\ref{tab:LECs}, most of them are pinned down in processes other than weak pion production. The so-called scale-independent parameter $\bar{\ell}_6$ was extracted from the electromagnetic charge radius of the pion $\langle r^2\rangle_\pi$ at $\mathcal{O}(p^4)$ in Ref.~\cite{Gasser:1983yg}. The value of $\ell_6$ at the renormalization scale $\mu$, denoted by $\ell_6^r$ in Eq.~\eqref{eq:UVshift}, can be obtained through the following renormalization group equation~\cite{Gasser:1983yg}
\bea
\ell_6^r=\frac{\beta_{\ell_6}}{16\pi^2}\bigg[\bar{\ell}_6+\ln\frac{M^2}{\mu^2}\bigg]\ ,
\eea
with $M^2=B_0(m_u+m_d)\simeq M_\pi^2$, where $B_0$ is a constant related to the quark condensate and $\beta_{\ell_6}=-1/6$ as can be seen from Eq.~\eqref{eq:betaell}. As usual in BChPT, we set $\mu=m_N$, which yields
\bea
\ell_6^r=(-1.34\pm1.74)\times 10^{-2}\ .
\eea

LECs $\tilde{c}_i$'s and $d_j^r$'s displayed in Table~\ref{tab:LECs}, except $\tilde{c}_6$ and $d^r_{22}$, have been fixed in pion-nucleon scattering, calculated up to $\mathcal{O}(p^3)$ using the $\delta$-counting within the EOMS scheme~\cite{Alarcon:2012kn}. This is exactly the same  approach employed in the present study. The model was fitted to the experimental phase shifts from Ref.~\cite{Arndt:2006bf}. On the other hand, the value of $\tilde{c}_6$ has been obtained in Ref.~\cite{Bauer:2012pv} by adjusting the corresponding chiral results for the magnetic moments of protons and neutrons, $\mu_p$ and $\mu_n$, to their empirical values from Ref.~\cite{Patrignani:2016xqp}. There are two determinations of this parameter in Ref.~\cite{Bauer:2012pv}: one is obtained without and the other with explicit $\Delta$'s which are present only in loops. We have chosen the former determination as the central value for $\tilde{c}_6$, since in the adopted power-counting rule loops with internal $\Delta$'s are of higher order and beyond our consideration. The difference between the two determinations is then assigned to the error of $\tilde{c}_6$. Specifically, we have $\tilde{c}_6=(1.35\pm0.04)~\text{GeV}^{-1}$ in the end\footnote{In Ref.~\cite{Bauer:2012pv}, the $\rho$ meson is explicitly included  in the calculation and the combination $\bar{c}_6=c_6+c_6^\rho$ is determined, where $c_6^\rho=-G_\rho/(2g_\rho)$ is the part saturated by the $\rho$, given in terms of parameters $G_\rho$, $g_\rho$,  related to $\rho$ interactions. In our case, without explicit $\rho$ meson, this $\rho$ contribution is absorbed by the LEC. Therefore, we identify our $\tilde{c}_6$ with  $\bar{c}_6$ rather than $c_6$. }. As for $d^r_{22}$, it is pinned down in  the extraction of the nucleon axial charge and radius from lattice QCD results in Ref.~\cite{Yao:2017fym}. Similarly to $\tilde{c}_6$, the $\Delta$ resonance is involved in the axial form factor only at loop level, hence we employ its value from the $\Delta$-less fit therein.

Finally, as demonstrated in Ref.~\cite{Bernard:2012hb}, the electromagnetic width of the $\Delta$ resonance can be expressed in terms of  the NLO $\pi N\Delta$ coupling $b_1$. Given that $\Gamma^{\rm em}/(\Gamma^{\rm em}+\Gamma^{\rm str})=0.55-0.65\%$ with $\Gamma_\Delta^{\rm str.}=118\pm 2$~MeV~\cite{Patrignani:2016xqp}, the value of $b_1$ is fixed to be $b_1=(4.98\pm0.27)/m_N$.  As mentioned in Ref.~\cite{Bernard:2012hb}, the sign of $b_1$ remains undetermined, but here we have chosen a positive sign as further discussed in the next subsection.

Apart from the known parameters discussed above,  there are still 7 unknown LECs: $d_1^r$, $d_6^r$, $d_8^r$, $d_{14}^r$, $d_{20}^r$, $d_{21}^r$ and $d_{23}^r$.\footnote{Some of these LECs, $d_8$, $d_9$, $d_{20}$ and $d_{21}$, also appear in  pion electroproduction on the nucleon. Their values have been determined in the analysis of that process in Ref.~\cite{Hilt:2013fda}. Although Ref.~\cite{Hilt:2013fda} uses the EOMS scheme, we cannot use their results directly because the $\Delta$ is not included in the calculation.} In our numerical computation, we assume them to be of natural size, namely
$d^r_{j}=0.0\pm 1.0~{\rm GeV}^{-2}$ with $ j \in\{ 1,6,8,14,20,21,23\}$. In view of the values of the known $d_j$'s in Table~\ref{tab:LECs}, our assumption seems quite reasonable. Note that the $d_2^r$ and $d_{15}^r$ can be obtained from $d_{1+2}^r$ and $d_{14-15}^r$ in Table~\ref{tab:LECs} with the help of the assumed $d_1^r$ and $d_{14}^r$ values, while the errors are propagated in quadrature.

\subsection{Total cross sections\label{sec:CS}}

Once the parameters in the hadronic transition amplitudes have been specified, we are now in the position to make predictions for experimental observables.  First, the (anti)neutrino-induced pion-production cross sections are calculated up to  $\mathcal{O}(p^3)$. The convergence properties of our results are then discussed.  We consider the muon flavor, for which the available measurements have been performed.
As previously explained in subsection~\ref{sec:ChPTandPC}, we expect our model to be reliable up to energies $E_\nu\simeq 415$~MeV, so that we are relatively far from the $\Delta$ pole and the $\delta$ counting is appropriate.

\begin{figure*}[ht]
\begin{center}
\epsfig{file=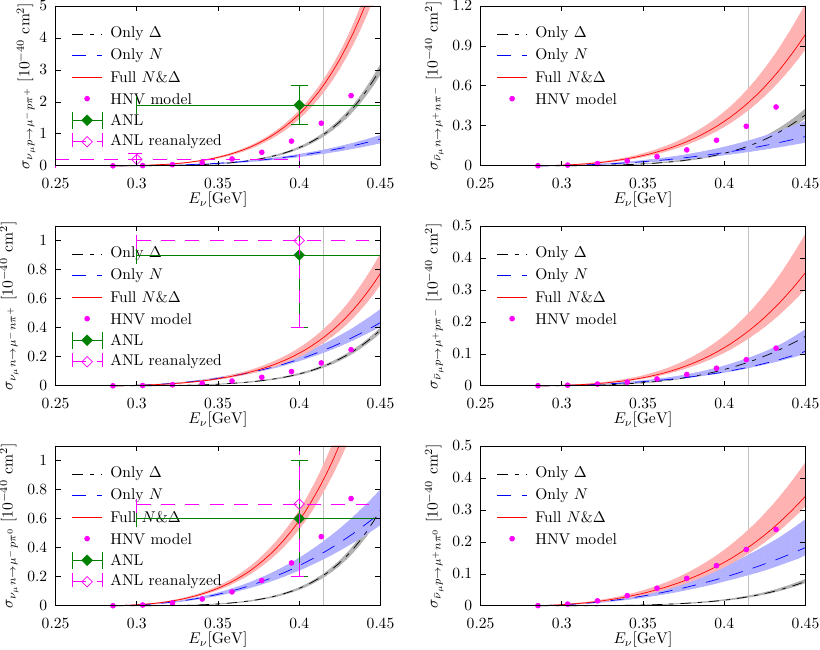,width=\textwidth}
\caption{Cross sections for weak pion production. The grey vertical line corresponds to $E_\nu=0.415$~GeV.
Dots correspond to the HNV model~\cite{Hernandez:2007qq}. The original ANL data are taken from Refs~\cite{Radecky:1981fn,Barish:1978pj}, while the recently reanalyzed ones are from Refs~\cite{Wilkinson:2014yfa,Rodrigues:2016xjj}. }
\label{fig:cs.all}
\end{center}
\end{figure*}

\begin{figure*}[t]
\begin{center}
\epsfig{file=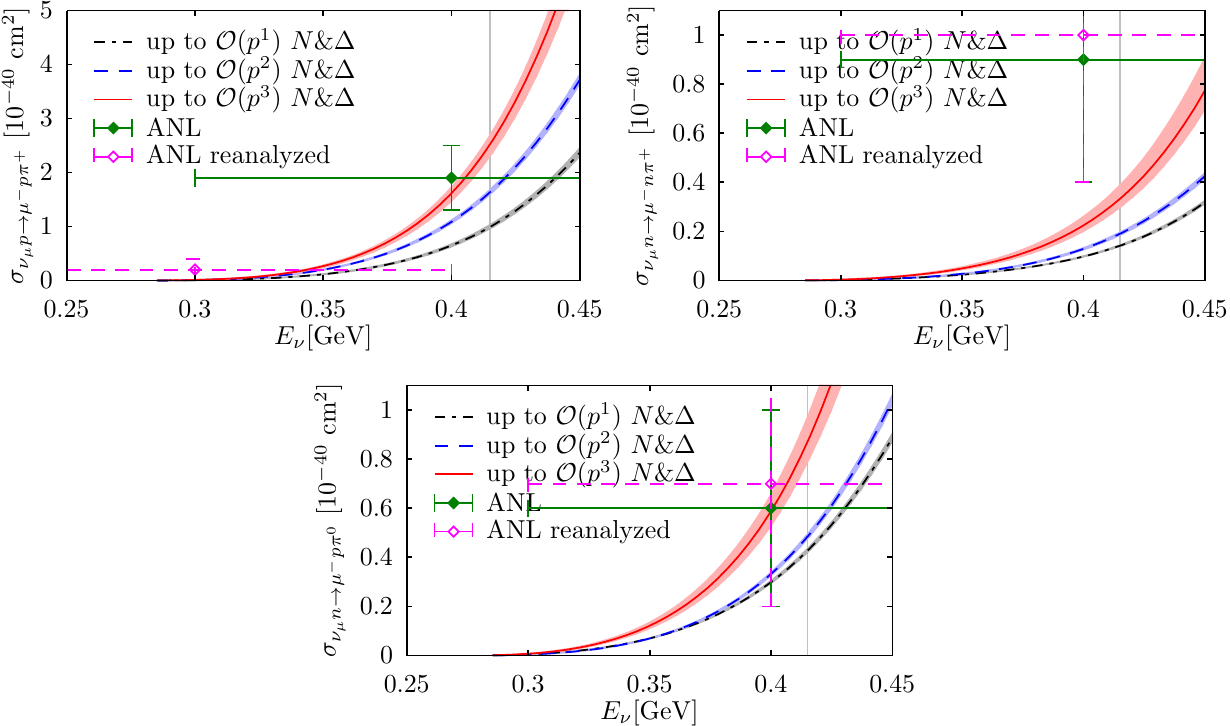,width=\textwidth}
\caption{Total cross sections for neutrino-induced pion production at different chiral orders. The grey vertical line corresponds to $E_\nu=415$~MeV. Original ANL data are taken from Refs~\cite{Radecky:1981fn,Barish:1978pj}, while the reanalyzed ones are from Refs~\cite{Wilkinson:2014yfa,Rodrigues:2016xjj}. }
\label{fig:cs.con}
\end{center}
\end{figure*}

  In the left (right) column of Fig.~\ref{fig:cs.all}, the results are shown for neutrino (antineutrino)-induced pion production, respectively.  The plots are displayed up to $E_\nu= 450$~MeV, slightly above our validity limit, to better show the trends of the curves. {The $\Delta$-width effect is taken into account as well by means of Eq~\eqref{eq:deltawidth}, though its contribution is of higher order.
Furthermore, its effects are really minor in the energy region we are concerned with. Its implementation enables us to eventually extend our results smoothly to higher energies, even passing the $\Delta$-peak}. Due to the nearby existence of the $\Delta$ pole, the $\Delta$ contribution (black dash-dotted line) increases rapidly in the region above $E_\nu\simeq 415$~MeV, as can be observed especially from the plot for the reaction $\nu_\mu p\to \mu^- p\pi^+$. Meanwhile, except for this latter channel, the nucleonic contribution (blue dashed line) grows steadily and dominates the total cross sections in the region below $E_\nu\simeq 415$~MeV.  The bands in the plots show the uncertainty associated to the error estimations of the LECs discussed in the previous section.

In the considered energy region,  there is only one experimental data point from the ANL measurements~\cite{Radecky:1981fn,Barish:1978pj,Rodrigues:2016xjj} for each neutrino-induced reaction channel. As can be seen in Fig.~\ref{fig:cs.all}, our full chiral predictions (red lines with bands), at $E_\nu\sim 400$~MeV, are in good agreement with the ANL data in the channels of $\nu_\mu p\to \mu^- p\pi^+$ and $\nu_\mu n\to \mu^-p\pi^0$. However, the theoretical cross section for the $\nu_\mu n\to \mu^- n\pi^+$ reaction is smaller than the central value of the experiment. Nevertheless, the chiral calculation for this latter channel is still consistent  with data  due to the large  experimental uncertainties. Unfortunately, for the antineutrino processes, so far there are no available data at low energies, preventing us from assessing our predictions. 

We also compare our results with those of the HNV model~\cite{Hernandez:2007qq}, which allows for a simple but meaningful comparison: the HNV phenomenological model, gives a good description of the weak pion production process for a wider range of neutrino energies well above 1 GeV. This model incorporates both the contributions from the $\Delta$ pole mechanisms and non-resonant terms constrained by chiral symmetry and given by the tree diagrams of Fig.~\ref{fig:treetopo} at their lowest order.  The counterparts of those diagrams in ChPT are represented by the LO $\Delta$-less tree diagrams of $\mathcal{O}(p)$ and  the $\Delta$-exchange ones of $\mathcal{O}(p^{3/2})$ and $\mathcal{O}(p^{5/2})$. In particular, for the $\Delta$ contribution, we find the following correspondence
\bea
C_{5A}(0) = \sqrt{\frac{2}{3}}h_A\ ,\qquad C_{3V}(0) =\frac{b_1m_N}{\sqrt{6}}\ ,
\eea
where $C_{5A}$ and $C_{3V}$ are two of the  Adler $N\to\Delta$ axial and vector form factors that are conventionally used in the literature~\cite{LlewellynSmith:1971uhs,Schreiner:1973mj}  including the HNV model. Imposing the values of $h_A$ and $b_1$ specified in the above subsection, we obtain $C_{5A}(0) \simeq 1.17\pm0.02$ and $C_{3V}(0) \simeq 2.01\pm 0.11$, which are comparable to the values $C_{5A}(0) = 1.2$,
and $C_{3V}(0) =2.13$ used in the HNV model and 
taken there from Refs.~\cite{Paschos:2003qr} and \cite{Lalakulich:2006sw} respectively. 
The small numerical difference in $C_{5A}(0)$  comes from a different choice of the $\Delta$ width in Ref.~\cite{Paschos:2003qr}. This observation also supports the choice of a positive $b_1$.
Note, that while HNV does not obey a systematic power counting neither includes loop diagrams,
some higher-order corrections are implemented through phenomenological form factors for the vertices in the axial and vector weak currents\footnote{In  particular, some additional higher order $\Delta$ couplings such as $C_{4A}$, $C_{4V}$ or  $C_{5V}$ are present. We do not consider them here as they would appear, together with many other contributions, in a higher order calculation.}. 
Our results are systematically larger than the HNV ones. This is mainly due to the inclusion of the $\mathcal{O}(p^3)$ terms coming both from tree and loop diagrams.  The enhancement improves the agreement with data  though the large error bars preclude any strong claim. Particularly interesting is the 
 $\nu_\mu n\to \mu^- n\pi^+$ channel where there is a large contribution of the $\mathcal{O}(p^3)$ terms but the results are still below data. 

In Fig.~\ref{fig:cs.con}, we display the total cross sections for the neutrino reactions order by order, in order to show the convergence properties of the chiral series\footnote{The same behavior is present in the case of the antineutrino reactions because neutrinos and antineutrinos share the same hadronic transition amplitude.}. For all the channels, a calculation with a higher chiral order brings the predictions closer to the experimental data. Moreover, the resulting contribution when stepping from $\mathcal{O}(p^2)$ up to $\mathcal{O}(p^3)$ is quite significant in the improvement of the predictions. On the other hand, it seems clear that next order effects could still be relevant. In fact, it has been shown for $\nu_\mu n\to \mu^- n\pi^+$, that the failure on the description of the ANL data might be cured by partially restoring  unitarity~\cite{Alvarez-Ruso:2015eva}. This can be  approximately done by imposing Watson's theorem for the dominant vector and axial multipoles~\cite{Alvarez-Ruso:2015eva}. In a systematic ChPT calculation, this corresponds to the inclusion of higher-order loops: especially those whose internal pion and nucleon lines can be put on shell. Another possible solution has been suggested in Ref.~\cite{Hernandez:2016yfb} and amounts to the need of extra higher-order contact terms.

\subsection{Pion angular distributions\label{sec:dis}}

Although for weak pion production differential cross sections are only available in averages over broad spectra of incoming neutrino energies, the low-energy predictions of the present approach may nonetheless be valuable for the comparison with future data and as benchmark for phenomenological models. Here, we discuss pion angular distributions in the so called isobaric frame, i.e. the CM frame of the outgoing $\pi N$ pair, usually considered for pion electroproduction, see e.g. Ref.~\cite{Alder:1972di}. To this end, the pion polar angle $\theta_\pi^\ast$ is defined with respect to the virtual $W$ boson direction $\hat{k}^\ast=(\vec{k}_1^\ast-\vec{k}_2^\ast)/|\vec{k}_1^\ast-\vec{k}_2^\ast|$, where the asterisk denotes a quantity in the $\pi N$ pair CM frame. Besides, the pion azimuthal angle $\phi_\pi^\ast$ is the angle between the reaction plane spanned by $\vec{k}_1^\ast$, $\vec{k}_2^\ast$ and the production plane, by $\vec{q}\,^\ast$, $\hat{k}^\ast$.

\begin{figure*}[h!]
\begin{center}
\epsfig{file=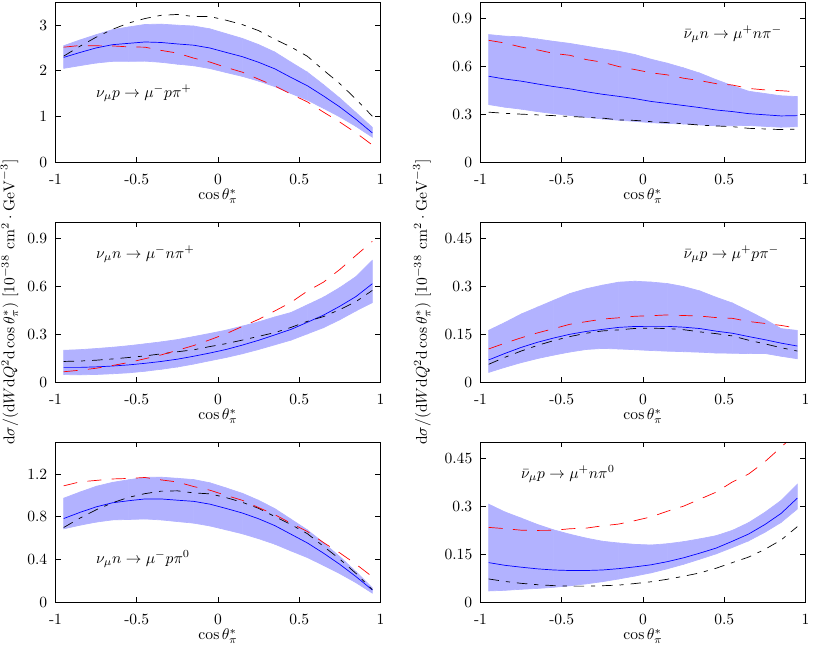,width=0.95\textwidth}
\caption{Pion polar-angle distributions for three different kinematic configurations.  
Dashed line for case (a): $E_\nu=0.315$ GeV, $W=1.11$ GeV, $Q^2=0.04$ GeV$^2$. 
Solid line and uncertainty band for case (b): $E_\nu=0.365$ GeV, $W=1.13$ GeV, $Q^2=0.05$ GeV$^2$.
Dash-dotted line for case (c): $E_\nu=0.415$ GeV, $W=1.15$ GeV, $Q^2=0.08$ GeV$^2$.
For a better visualization, dashed and dash-dotted lines have been multiplied by $15$ and $1/3$ respectively.}
\label{fig:ct}
\end{center}
\end{figure*}
\begin{figure*}[h!]
\begin{center}
\epsfig{file=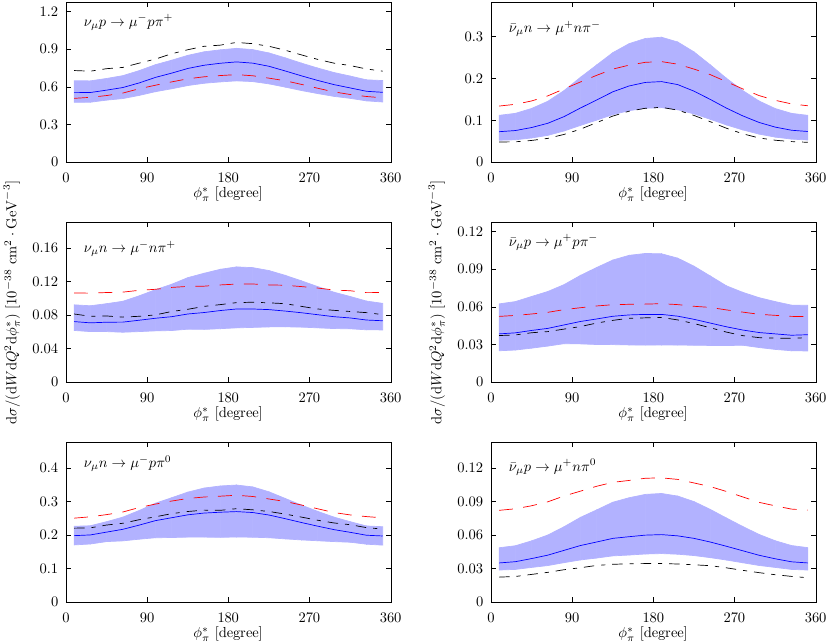,width=0.9\textwidth}
\caption{Pion azimuthal angular distributions. Same definitions as in Fig.~\ref{fig:ct}.}
\label{fig:phi}
\end{center}
\end{figure*}

Numerical results for pion polar and azimuthal angular distributions, ${\rm d}\sigma/({\rm d}W{\rm d}Q^2{\rm d}\cos\theta_\pi^\ast)$ and ${\rm d}\sigma/({\rm d}W{\rm d}Q^2{\rm d}\phi^\ast_\pi)$, are shown in Figs.~\ref{fig:ct} and \ref{fig:phi}, respectively. Three different sets of $\{E_\nu,W,Q^2\}$ inside the adopted validity region  have been chosen: (a) close to threshold, (b) at at intermediate $E_\nu$ value, and (c) at the upper neutrino-energy limit. In Figs.~\ref{fig:ct} and \ref{fig:phi}, to render the comparison easy, the results for case (a) and case (c) have been scaled by factors of $15$ and $1/3$, respectively. 
It can be observed that the shapes of the $\cos\theta^\ast_\pi$ distribution for the three different cases in each channel are similar but there are differences among channels. This observation also holds true for the azimuthal $\phi^\ast_\pi$ distributions.  One can also see from Fig.~\ref{fig:phi} that the $\phi_\pi^\ast$ distributions are almost symmetric around $\pi$,  indicating that the asymmetries proportional to $\sin\phi_\pi^\ast$ and $\sin2\phi_\pi^\ast$ identified in Ref.~\cite{Hernandez:2007qq}, are negligible at low energies. Representatively, for case (b) we display the error bands resulting from the propagation of the LEC uncertainties. 

\subsection{Multipole expansion\label{sec:mul}}

Multipole amplitudes carry detailed information about the hadronic transition induced by the weak interaction. The formalism for the multipole expansion of the hadronic matrix elements was developed in detail in Ref.~\cite{Adler:1968tw}, thus here we only show the formulae needed to establish the connection to our chiral amplitudes. Based on Ref.~\cite{Adler:1968tw} we can write (for any $\epsilon^\mu\propto L^\mu$)
\bea \label{eq:isobaricamp}
\epsilon^\mu H^{V (\pm)}_\mu=-i\sum_{j=1}^{6}\mathscr{F}_j^{V({\pm})}\eta_f^\dagger\Sigma_j^V\eta_i\ ,\qquad \epsilon^\mu H^{{A (\pm)}}_\mu=-i\sum_{j=1}^{8}\mathscr{G}_j^{A({\pm})}\eta_f^\dagger\Lambda_j^A\eta_i\ ,
\eea
where $H^{V,A(\pm)}$ stand for the second and first terms in Eq.~(\ref{eq:lordecom}), in this order; $\eta_i$ and $\eta_f$ are two-component Pauli spinors of the initial and the final nucleons in the isobaric frame. For the explicit expression of the Pauli operators, $\Sigma_j^V$ and $\Lambda_j^A$, we refer the reader to Ref.~\cite{Adler:1968tw}. The above equations allow us to relate the isobaric amplitudes $\mathscr{F}_j^{V}$ ($\mathscr{G}_j^{A}$) to $V_i$ ($A_i$). The resulting expressions are relegated to Appendix~\ref{sec:isobaramp}. It is convenient to introduce linear combinations of $H^{V,A(\pm)}_\mu$, which denote transitions to  pure isospin states of the final pion-nucleon pair
\bea
\label{eq:isospin}
H^{V,A(I=1/2)}_\mu = H^{V,A(+)}_\mu + 2 H^{V,A(-)}_\mu \,, \qquad 
H^{V,A(I=3/2)}_\mu = H^{V,A(+)}_\mu - H^{V,A(-)}_\mu \,.
\eea
Equivalent combinations for the isobaric amplitudes obviously apply. It is now possible to write multipole expansions of $\mathscr{F}_j^{V (I)}$ and $\mathscr{G}_j^{A (I)}$ for transitions to pion-nucleon states with angular momentum $\ell$. They are given in Ref.~\cite{Adler:1968tw}, as well as the corresponding inversion formulas. 

In general, for any given angular momentum $\ell$, there are six vector multipole amplitudes, $M_{l\pm}$, $E_{l\pm}$ and $L_{l\pm}$,  and eight axial-vector ones, $\mathcal{M}_{l\pm}$, $\mathcal{E}_{l\pm}$ and $\mathcal{L}_{l\pm}$ and $\mathcal{H}_{l\pm}$. Specifically, for $S$ ($\ell=0$) and $P$ ($\ell=1$) waves, there are only  5 and 12 amplitudes, respectively. As illustration, their values for both $I=\frac{1}{2}$ and $I=\frac{3}{2}$ are displayed in Table~\ref{tab:mps} at $W=1.13$~GeV and $Q^2=0.05$~GeV$^2$, which is a typical point of the available phase space in the energy range considered in this work,\footnote{For the purpose of benchmarking other theoretical models, multipole amplitudes at any other values of $W$ and $Q^2$ are available from the first author (D. Y.) upon request.}. The errors are propagated from the uncertainties of the involved LECs. In the case of the imaginary parts, they are negligible and, therefore, not shown. One can see that the $S$-wave multipoles in Table~\ref{tab:mps} are larger than the corresponding $P$-wave ones by one order of magnitude.\footnote{In Ref.~\cite{Bernard:1993xh}, the $S$-wave axial-vector multipole amplitudes are calculated using HB ChPT but only at threshold and in the approximation of zero lepton mass. Note also that those multipole amplitudes are obtained with a different normalization with respect to ours, and have dimensions of $[{\rm mass}]^{-1}$. }  
We have checked that the $P$-wave multipoles decrease rapidly to zero when $W$ goes to threshold.

\begin{table}[h]
\caption{$S$- and $P$-wave multipole amplitudes calculated at $W=1.13$~GeV and $Q^2=0.05$~GeV$^2$. Here, the multipole amplitudes are dimensionless by definition. }\label{tab:mps}
\vspace{-0.5cm}
\bea
\begin{array}{c|cc}
\hline\hline
&{ I=1/2}&{ I=3/2}\\
\hline
E_{0+}&(29.5^{+0.75}_{-0.91},4.79)&(-15.6^{+0.5}_{-0.5},1.20)\\
L_{0+}&(-197^{+75}_{-70},-32.1)&(188^{+35}_{-34},-6.58)\\
\mathscr{M}_{0+}&(7.25^{+0.51}_{-0.32},0.116)&(-1.03^{+0.43}_{-0.40},-0.219)\\
\mathscr{L}_{0+}&(7.14^{+24.3}_{-17.2},7.14)&(-67.2^{+9.9}_{-11.8},-0.219)\\
\mathscr{H}_{0+}&(8.90^{+18.9}_{-13.2},5.37)&(-48.6^{+7.6}_{-9.5},-0.125)\\
\hline
M_{1+}&(-7.85^{+1.52}_{-2.09},0.107)&(21.8^{+1.2}_{-2.3},1.36)\\
E_{1+}&(3.27^{+0.40}_{-0.47},-0.0812)&(-2.21^{+0.24}_{-0.20},-0.127)\\
L_{1+}&(-27.9^{+4.1}_{-3.5}, 0.612)&(24.4^{+1.9}_{-2.2},1.54)\\
M_{1-}&(-15.8^{+1.1}_{-2.1}, -0.500)&(-7.27^{+1.51}_{-2.12},0.147)\\
L_{1-}&(-50.9^{+10.2}_{-9.2},0.432)&(47.0^{+4.8}_{-5.2},3.36)\\
\mathscr{M}_{1+}&(157^{+7}_{-4},-0.796)\cdot10^{-3}&(-14.7^{+0.69}_{-0.39},-0.318)\cdot 10^{-2}\\
\mathscr{E}_{1+}&(1.42^{+1.96}_{-2.44},-0.120)&(-39.7^{+1.3}_{-1.3},-2.79)\\
\mathscr{L}_{1+}&(-1.09^{+2.45}_{-1.64},0.0744)&(20.9^{+1.0}_{-1.0},1.45)\\
\mathscr{H}_{1+}&(-0.641^{+1.560}_{-1.027},0.0469)&(12.7^{+0.7}_{-0.6},0.881)\\
\mathscr{E}_{1-}&(1.69^{+2.47}_{-2.34},1.68)&(-1.99^{+1.69}_{-2.10},0.140)\\
\mathscr{L}_{1-}&(4.74^{+4.17}_{-4.36},2.12)&(-5.17^{+3.04}_{-4.02},0.121)\\
\mathscr{H}_{1-}&(3.52^{+3.09}_{-3.59},1.34)&(-3.40^{+2.03}_{-2.68},0.0683)\\
\hline\hline
\end{array}\nonumber
\eea
\end{table}

\section{Summary and outlook\label{sec:summary}}
Charged current (anti)neutrino-induced pion production off the nucleon at low energies has been systematically studied  for the first time within the framework of manifestly relativistic baryon chiral perturbation theory up to $\mathcal{O}(p^3)$, (NNLO), for the low-energy chiral representation of the hadronic-transition amplitude. The $\Delta$(1232) resonance has been included explicitly using the $\delta$-counting rule. To tackle the  power-counting violation of the nucleon loops we have performed the renormalization in the EOMS approach~\cite{Gegelia:1999gf,Gegelia:1999qt,Fuchs:2003qc} in which the power counting is restored by means of
finite shifts of the LEC values in the chiral effective Lagrangians  after the conventional UV subtraction in the $\overline{\rm MS}$-1 scheme.

Remarkably, at this order, most of the involved LECs (15 out of 22) have been previously determined in other processes such as pion-nucleon scattering. Furthermore, another 4 of the remaining unknown LECs in the $\mathcal{O}(p^3)$ $\pi N$ Lagrangian may be obtained in the future from available pion electroproduction data. For numerical estimates, the unknown LECs have been  assumed to be of natural size. Consequently, we have predicted the total cross sections in all the physical reaction channels, both for neutrino- and antineutrino-induced pion production. We have also estimated the theoretical uncertainties due to the limited knowledge of some LECs. Our results are expected to be reliable up to the neutrino laboratory energy of $E_\nu=415$~MeV, which is relatively close to the threshold and well below the $\Delta$ peak. Hence, the energy range is well suited for the adopted $\delta$-counting. Nonetheless, mechanisms involving the $\Delta$ resonance contribute significantly to all production  channels, especially to the $\nu_\mu p\to \mu^- p\pi^+$ one.

It has been found that our predictions are consistent with the few existing experimental ANL data for the neutrino-induced processes except $\nu_\mu n\to \mu^-n\pi^+$. This might indicate that higher-order contributions are still relevant for this channel as suggested by the more phenomenological study of Ref.~\cite{Hernandez:2016yfb}. Lacking a full calculation, such higher-order contributions might be approximated by unitarity corrections or some extra contact counterterms. So far there are no low-energy experimental data for antineutrino-induced pion production on nucleons. Our results for these processes provide a set of theoretical predictions that fully rely on ChPT.

Finally, our chiral representation of weak pion production can be applied to study various low-energy theorems in the future. It can also be adapted to make a comprehensive analysis of pion photo-, electro-production and neutral-current induced weak production in all physical channels by further incorporating the isoscalar vector part of the hadronic currents. Most importantly, the present study provides a well founded low-energy benchmark for phenomenological models aimed at the description of weak pion production processes in the broad kinematic range of interest for current and future neutrino-oscillation experiments.

\acknowledgments
This research has been supported by the Spanish Ministerio de Econom\'ia y Competitividad (MINECO) and the European Regional Development Fund (ERDF), under contracts FIS2014-51948-C2-1-P, FIS2014-51948-C2-2-P, FIS2017-84038-C2-1-P, FIS2017-84038-C2-2-P, SEV-2014-0398, and by Generalitat Valenciana under contract PROMETEOII/2014/0068. It has also been supported by the Deutsche Forschungsgemeinschaft (DFG).

\appendix

\begin{widetext}
\section{Chiral hadronic amplitudes at tree level \label{sec:tree.cons}}

In what follows, all the tree amplitudes corresponding to the diagrams specified in Table~\ref{tab:tree} are listed.  
We use the abbreviations: 
\bea
\Sigma_{x}&=x-\mnuc^2+\mpir^2, \quad \Sigma_{x}^\prime=x-\mnuc^2+t_1\ , \\  \Delta_{x}&=x-\mnuc^2-\mpir^2\ , \quad  \Delta_{x}^\prime=x-\mnuc^2-t_1\ ,
\eea
with $x\in\{s_2, u\}$. The Mandelstam variable $u$ is defined as
$
u\equiv(p_1-q)^2
$,
and hence can be written in terms of the variables in Eq.~\eqref{eq:mans} via
$
u= 2m_N^2-s_2-t-t_1
$.
Hereafter, the Lorentz indices of the axial and vector operators are suppressed. Furthermore, we shall use the shorthands: 
\bea
O^{A,V}_{i\pm j\pm k\pm \cdots}=O^{A,V}_{i}\pm O^{A,V}_{j}\pm O^{A,V}_{k}\pm\cdots\ ,i,j,k\in\{1,\cdots,8\}\,.
\eea

As for the $\Delta$-exchange diagrams, the $\Delta$-width effect can be included through the following substitution:
\bea\label{eq:deltawidth}
\frac{1}{m_\Delta^2-s_\Delta}\to \frac{1}{m_\Delta^2-i\,m_\Delta \Gamma_\Delta(s_\Delta)-s_\Delta}\ ,\quad s_\Delta\equiv p_\Delta^2\ ,
\eea 
with the energy-dependent width given by~\cite{Gegelia:2016pjm}
\bea
\Gamma_\Delta(s_\Delta)=\frac{h_A^2\lambda^{\frac{3}{2}}(s_\Delta,M_\pi^2,m_N^2)}{192\pi F_\pi^2s_\Delta^3}\big[(s_\Delta-M_\pi^2+m_N^2)m_\Delta+2s_\Delta\,m_N\big] \theta(\sqrt{s_\Delta }-m_N - M_\pi) \ ,
\eea
being  $\lambda(a,b,c)\equiv a^2+b^2+c^2-2ab-2ac-2bc$ the K\"all\'en function and $\theta(x)$ the step function.

\begin{table*}[ht]
\caption{Labels for tree diagrams.
}\label{tab:tree}
\vspace{-0.5cm}
\bea
\begin{array}{cc|ccc|cc}
\hline\hline
&&\multicolumn{3}{c|}{\Delta\text{-less diagram}} &\multicolumn{2}{c}{\Delta\text{-exchange diagram}}\\
\textbf{Topology}&\textbf{Label}&\mathcal{O}(p)&\mathcal{O}(p^2)&\mathcal{O}(p^3)&\mathcal{O}(p^{3/2})&\mathcal{O}(p^{5/2})\\
\hline
\text{Type (a)}&T^{A(\Delta)}_{ij}& T^A_{11}&T^A_{21}& T^A_{31}, \quad T^A_{13}& T^{A\Delta}_{11}&T^{A\Delta}_{21}\\
\text{Type (b)}&T^{B(\Delta)}_k& T^B_{1}&T^B_{2}& T^B_{3}&&\\
\text{Type (c)}&T^{C(\Delta)}_{ij}& T^C_{11}&T^C_{12}& T^C_{13}, \quad T^C_{31}& T^{C\Delta}_{11}&T^{C\Delta}_{12}\\\
\text{Type (d)}&T^{D(\Delta)}_{im}& T^D_{12}&& T^D_{14},\quad T^D_{32}\\
\text{Type (e)}&T^{E(\Delta)}_{ijm}& T^E_{112}&& T^E_{114}, \quad T^E_{132},\quad T^E_{312}& T^{E\Delta}_{112}&\\
\text{Type (f)}&T^{F(\Delta)}_{km}& T^F_{12}&T^F_{22}& T^F_{32},\quad T^F_{14}\\
\text{Type (g)}&T^{G(\Delta)}_{ijm}& T^G_{112}&& T^G_{114}, \quad T^G_{312},\quad T^G_{132}& T^{G\Delta}_{112}&\\
\hline\hline
\end{array}\nonumber
\eea
\end{table*}

In the following $H^\pm_\mu = -2 \sqrt{2} \,\mathcal{H}^\pm_\mu$.

\subsection{At \texorpdfstring{$\mathcal{O}(p)$:}{O(p):}}
\begin{itemize}
\item {Diagram $T^A_{11}$:}
\bea
\mathcal{H}^\pm_\mu&=&\frac{g}{8F(m^2-s_2)}\bigg\{g\big[(m+m_N)(2\MOA_{1}-\MOA_{7})+(\mnuc^2-s_2)\MOA_{8}\big]\no
&-&\big[(m+m_N)(2\MOV_{1}-\MOV_{7})+(\mnuc^2-s_2)\MOV_{8}\big]\bigg\}\ .
\eea
\item {Diagram $T^B_1$:}
\bea
\mathcal{H}^+_\mu=0\ ,\qquad
\mathcal{H}^-_{\mu}=-\frac{1}{4F}\bigg\{\MOA_{8}-g\,\MOV_{8}\bigg\}\ .
\eea
\item {Diagram $T^C_{11}$: }
\bea
\mathcal{H}^+_\mu&=&\frac{g}{8F(m^2-u)}\bigg\{g\big[(m+m_N)\MOA_{7}-(m_N^2-u)\MOA_{8}\big]\no
&+&\big[(m+m_N)\MOV_{7}+(m_N^2-u)\MOV_{8}\big]
\bigg\}=-
\mathcal{H}^-_\mu\ .
\eea
\item {Diagram $T^D_{12}$:}
\bea
\mathcal{H}^+_{\mu}=0\ ,\qquad 
\mathcal{H}^-_{\mu}=\frac{g~m_N\,\MOV_{1+2-3}}{2F(M^2-t)}.
\eea
\item {Diagram $T^E_{112}$:}
\bea
\mathcal{H}^\pm_\mu&=&\frac{g^2}{8F(t_1-M^2)(m^2-s_2)}\bigg\{2m_N(m_N^2-s_2)\MOA_{1-2+3}\no
&+&(3m_N^2+s_2)\MOA_{4-5+6}\bigg\}\ .
\eea
\item {Diagram $T^F_{12}$:}
\bea
\mathcal{H}^+_{\mu}=0\ ,\qquad 
\mathcal{H}^-_{\mu}=-\frac{\MOA_{4-5+6}}{4F(M^2-t_1)}\ .
\eea
\item {Diagram $T^G_{112}$:}
\bea
\mathcal{H}^+_\mu&=&\frac{g^2}{8F(t_1-M^2)(m^2-u)}\bigg\{2m_N(m_N^2-u)\MOA_{1-2+3}\no
&-&(3m_N^2+u)\MOA_{4-5+6}\bigg\}=-
\mathcal{H}^-_\mu\ .
\eea
\end{itemize}

\subsection{At \texorpdfstring{$\mathcal{O}(p^2)$:}{O(p2):}}
\begin{itemize}
\item {Diagram $T^A_{21}$:}
\bea
\mathcal{H}^\pm_\mu&=&\frac{c_6\, \g}{16 \F \mnuc (s_2-\mnuc^2)}\bigg\{
\mnuc^2 (7 \MOV_{1}+\MOV_{2+3}-3 \MOV_{7})\no
&+&2 \mnuc\big[\MOV_{4+5+6}+(\mnuc^2-s_2) \MOV_8\big]+s_2\,\MOV_{1-2-3-7}
\bigg\} .
\eea
\item {Diagram $T^B_{2}$:}
\bea
\mathcal{H}^+_\mu &=& \frac{1}{4\mnuc^2\F}\bigg\{c_2 \big[\Delta_u\MOA_2\-\Delta_{s_2}\MOA_3\big]-4\,c_3 \mnuc^2\MOA_1\bigg\}\ ,\no
\mathcal{H}^-_\mu &=&\frac{1}{8\F\mnuc}\bigg\{4c_4\mnuc\MOA_{1-7}+c_6  (\MOA_{1-2-3-7}+2 \mnuc \,\MOA_8)\bigg\} .
\eea
\item {Diagram $T^C_{12}$:}
\bea
\mathcal{H}^+_\mu&=&-H^-_\mu=\frac{c_6\, \g}{16 \F \mnuc (\mnuc^2-u)}\bigg\{
\mnuc^2 (\MOV_{1-2-3}+3 \MOV_7)\no
&+&2 \mnuc(\mnuc^2-u) \MOV_8
-u\,\MOV_{1-2-3-7}+2\mnuc\MOV_{4-5-6}
\bigg\} \ .
\eea
\item {Diagram $T^F_{22}$:}
\bea
\mathcal{H}^+_\mu&=&\frac{\MOA_{1-2+3}}{8 \F \mnuc^2 (\mpir^2-t_1)}\bigg\{  16 c_1 \mnuc^2 \mpir^2-4 c_3 \mnuc^2 (\mpir^2-t+t_1)\no
&&-c_2 \big[(s_2+u)(t_1+\mpir^2)-2(s_2 u+\mpir^2 t_1)+2\mnuc^2(\mnuc^2-t)\big]
\bigg\}\ ,\no
\mathcal{H}^-_\mu&=&\frac{c_4}{4 \F (\mpir^2-t_1)}  \bigg\{(s_2-u) \MOA_{1-2+3}-4 \mnuc \MOA_{4-5+6}\bigg\}\ .
\eea
\end{itemize}

\subsection{At \texorpdfstring{$\mathcal{O}(p^3)$:}{O(p3):}}
\begin{itemize}
\item{Diagram $T^{A}_{31}$:}
\bea
\mathcal{H}^\pm_\mu&=&-\frac{\g}{8 \F \mnuc (\mnuc^2-s_2)} \bigg\{4 d_{16} \mnuc \mpir^2 \big[-2 \mnuc (2 \MOA_1-\MOA_7)+\MOA_8 (s_2-\mnuc^2)\big]\no
&&+ d_6 \big[2 \big(2 \mnuc \MOV_5+\MOV_2 (\mnuc^2-s_2)\big) t_1-\Dstpr \big(2 \mnuc \MOV_{4-5+6}+\MOV_{1-2+3} (\mnuc^2-s_2)\big)\big]\no
&&- d_{22} \mnuc \big[\MOA_{4-5+6} (3 \mnuc^2+s_2)+2 \mnuc (2 \MOA_1-\MOA_7) t_1\no
&&+(\mnuc^2-s_2) (2 \mnuc \MOA_{1-2+3}+\MOA_8 t_1)\big]\bigg\}.
\eea
\item{Diagram $T^{A}_{13}$:}
\bea
\mathcal{H}^\pm_\mu&=&-\frac{ (d_{18}-2 d_{16} ) \mpir^2}{4\F(\mnuc^2-s)}\bigg\{\g\big[2m_N(2\MOA_{1}-\MOA_{7})+(\mnuc^2-s_2)\MOA_{8}\big]\no
&-&\big[2m_N(2\MOV_{1}-\MOV_{7})+(\mnuc^2-s_2)\MOV_{8}\big]\bigg\}\ .
\eea
\item{Diagram $T^{B}_{3}$:}
\bea
\mathcal{H}^+_\mu&=&\frac{d_8}{\F \mnuc}\bigg\{\Dst \MOV_2-2 \mnuc \MOV_{5+6}-\MOV_{1-7} (4 \mnuc^2-t)-\MOV_3 (\Dstpr+t)\nonumber\\
&&+\mnuc \MOV_8 (\Dst+\Dstpr+t)\bigg\}+\frac{d_{14} (s_2-u)}{4 \F \mnuc}\MOA_{1-7}\no
&&+\frac{d_{15}}{8 \F \mnuc}\bigg\{4 \mnuc \MOA_{5+6}-\MOA_{1+2+3-7} (\Dst+\Dstpr+t)\bigg\}\no
&&-\frac{d_{23}}{2\F}\bigg\{2 \MOA_{5+6}-\MOA_8 (\Dst+\Dstpr+t)\bigg\}\ ,\no
\mathcal{H}^-_\mu&=&-\frac{d_1 (s_2-u)}{2 \F \mnuc}\MOA_1 +\frac{d_2}{4 \F \mnuc}\bigg\{\MOA_1 (u-s_2)-\MOA_{2+3} (\mpir^2-t+t_1)\bigg\}\no
&&-\frac{d_3}{12 \F \mnuc^3}\bigg\{ \Dst (2 \Dst+t)\MOA_3+ (2 \Dstpr+t) (\Dstpr+t)\MOA_2\bigg\}-\frac{d_5\mpir^2}{\F \mnuc} \MOA_{2+3}\no
&&+\frac{d_6}{4 \F \mnuc}\big[(\Dst+t)\MOA_2 -\Dstpr \MOA_3 \big]+\frac{(2 d_{16}-d_{18}) \mpir^2 \MOV_8}{2 \F}\no
&&-\frac{d_{20}}{16 \F \mnuc^2}\bigg\{2(2\mnuc \MOV_3+\MOV_6 ) \Dst+2(2\mnuc \MOV_2+\MOV_5) (\Dstpr+t)\no
&&-\MOV_8 ((\Dst+\Dstpr) (\Dst+\Dstpr+t+\mnuc^2)+2 \Dst (t_1-s_2)-\mnuc^2 (\mpir^2-t_1))\bigg\}\no
&&-\frac{d_{21}}{4\F}\big[4 \mnuc \MOV_1+2 \MOV_4-\MOV_8 (\mpir^2-t+t_1)\big]\no
&&+\frac{d_{22}}{8\F}\big[4 \mnuc \MOV_{2-3}+2 \MOV_{5-6}-\MOV_8 (\mpir^2-t-t_1)\big].
\eea
\item{Diagram $T^{C}_{13}$:}
\bea
\mathcal{H}_\mu^+&=&-\frac{\g}{8 \F \mnuc (\mnuc^2-u)} \bigg\{4 d_{16} \mnuc \mpir^2 \big[2 \mnuc \MOA_7-(\mnuc^2 - u)\MOA_8\big]\no
&&+d_{22} \mnuc \big[\mnuc (2 \MOA_7 t_1+\mnuc (2 \mnuc \MOA_{1-2+3}-3 \MOA_{4-5+6}-\MOA_8 t_1))\no
&&-(2 \mnuc \MOA_{1-2+3}+\MOA_{4-5+6}-\MOA_8 t_1) u\big]+d_6 \big[\mnuc^3 (\mnuc\MOV_{1-2+3}+2 \MOV_{4-5+6})\no
&&+\mnuc (\mnuc \MOV_{1-2-3}+2 \MOV_{4-5-6}) t_1-\big(2 \mnuc (\mnuc\MOV_{1-2+3}+ \MOV_{4-5+6})\no
&&+\MOV_{1-2-3} t_1\big) u+\MOV_{1-2+3} u^2\big]\bigg\}=-\mathcal{H}_\mu^-.
\eea
\item{Diagram $T^{C}_{31}$:}
\bea
\mathcal{H}^+_\mu&=&\frac{ (d_{18}-2 d_{16} ) \mpir^2}{4\F(\mnuc^2-u)}\bigg\{\g\big[2m_N\MOA_{7}-(m_N^2-u)\MOA_{8}\big]\no
&+&\big[2m_N\MOV_{7}+(m_N^2-u)\MOV_{8}\big]
\bigg\}=-
\mathcal{H}^-_\mu\ .
\eea
\item{Diagram $T^{D}_{14}$:}
\bea
\mathcal{H}_\mu^+&=&0\ ,\no
\mathcal{H}_\mu^-&=&\frac{\g  \mnuc }{2 \F^3 (\mpir^2-t)}\bigg\{l_6  (t-\mpir^2)\MOV_{1-2+3}+  \big(l_6\,t_1-2 l_6\, \mpir^2\big)\MOV_{1+2-3}\bigg\}\ .
\eea
\item{Diagram $T^{D}_{32}$:}
\bea
\mathcal{H}_\mu^+=0\ ,\qquad H_\mu^-=-\frac{(2 d_{16}-d_{18})  \mpir^2 \mnuc \MOV_{1+2-3}}{\F (\mpir^2-t)}\ .
\eea
\item{Diagram $T^{E}_{114}$+$T^{E}_{312}$+$T^{E}_{132}$:}
\bea
 \mathcal{H}_\mu^\pm&=&-
\frac{\big[2\g \F^2(2 d_{16} - d_{18} )+\g^2 l_4\big]  \mpir^2}{4 \F^3 (\mnuc^2-s_2) (\mpir^2-t_1)} \no
&&\times\bigg\{2 \mnuc (\mnuc^2-s_2)\MOA_{1-2+3} +(s_2+3\mnuc^2) \MOA_{4-5+6}\bigg\}.
\eea
\item{Diagram $T^{F}_{14}$:}
\bea
\mathcal{H}_\mu^+=0\ ,\qquad H_\mu^-=-\frac{l_4  \mpir^2 \MOA_{4-5+6}}{2 \F^3 (\mpir^2-t_1)}\ .
\eea
\item{Diagram $T^{F}_{32}$:}
\bea
\mathcal{H}_\mu^+&=&\frac{(d_{14}-d_{15})(u-s_2)}{8 \F \mnuc (\mpir^2-t_1)} \bigg\{4 \mnuc \MOA_{4-5+6}+(\Du+\Dupr+t)\MOA_{1-2+3} \bigg\} \ ,\no
\mathcal{H}_\mu^-&=&\frac{u-s_2}{16 \F \mnuc^3 (\mpir^2-t_1)}\MOA_{1-2+3}\bigg\{16 d_5 \mnuc^2 \mpir^2+4 (d_1+d_2) \mnuc^2 (\mpir^2-t+t_1)\no
&&+d_3\big[(2 s_2 (s_2 + t) - (2 s_2 + t) t_1 + 2 t_1^2 -
  2 \mnuc^2 (\Dst+\Dstpr  + t + m_N^2) \no
  &&- 
  \mpir^2 (-2 \mpir^2 + 2 s_2 + t + 2 t_1))\big]
\bigg\}.
\eea
\item{Diagram $T^{G}_{114}$+$T^{G}_{132}$+$T^{G}_{312}$:}
\bea
\mathcal{H}_\mu^+=-H_\mu^-&=&-\frac{\big[2\g \F^2(2 d_{16} - d_{18} )+\g^2 l_4\big]  \mpir^2}{(4 \F^3 (\mpir^2-t_1) (\mnuc^2-u))} \no
&&\times\bigg\{2 \mnuc(\mnuc^2-u)  \MOA_{1-2+3}-(3\mnuc^2+u) \MOA_{4-5+6}\bigg\}.
\eea
\end{itemize}
\subsection{At \texorpdfstring{$\mathcal{O}(p^{3/2})$:}{O(p3/2):}}

\begin{itemize}
\item {Diagram $T^{A\Delta}_{11}$:}
\bea
\mathcal{H}^+_\mu=-2H^-_\mu&=&\frac{\h^2}{18 \F \mdel^2 (\mdel^2 - s_2)}\bigg\{2 \mdel^3 \MOA_{1+7}
-2 \left(\mnuc \MOA_{1+3}+\MOA_{4+6}\right) \Sigma_{s_2}\no
&+&\mdel^2 \big[4 \MOA_4-2 \MOA_6+2 \mnuc \MOA_{1+7}+\MOA_8 \Delta_{s_2}\big]\no
&+&\mdel \big[\Sigma_{s_2} (2 \MOA_3+\MOA_7-\mnuc \MOA_8)+2\mpir^2 \MOA_{1+3}+2\mnuc  \MOA_{4+ 6}\big]
\bigg\} \ .
\eea

\item {Diagram $T^{C\Delta}_{11}$:} 
 \bea
 \mathcal{H}^+_\mu=2H^-_\mu&=&\frac{\h^2}{18 \F \mdel^2 (\mdel^2 - u)}\bigg\{
2 \mdel^3 (3 \MOA_1- \MOA_7)
-2 \left(\mnuc \MOA_{1 - 2} - \MOA_{4-5}\right) \Sigma_{u}\no
&+&\mdel^2\big[2\mnuc (3 \MOA_1-\MOA_7)-2 (2 \MOA_4+\MOA_5)-\MOA_8\Delta_u\big]\no
 &-&\mdel\big[-2 \mpir^2 \MOA_{1 - 2} + 2 \mnuc \MOA_{4 - 5}\no
 &&-(2 \MOA_{1 -2} - \MOA_7 + \mnuc \MOA_8) \Sigma_u\big]\bigg\}\ .
 \eea
\item {Diagram $T^{E\Delta}_{112}$:}  
\bea
\mathcal{H}^+_\mu=-2H^-_\mu&=&\frac{\h^2}{18 \F \mdel^2 (\mdel^2 - s_2)(t_1-\mpir^2)}\bigg\{\big[\mnuc \MOA_{1-2+3}+\MOA_{4-5+6} \big] \Sigma_{s_2}\Sigma_{s_2}^\prime\no
&&-\mdel^3 \big[4 \mnuc \MOA_{4-5+6}+\MOA_{1-2+3} (s_2+3 u-4 \mnuc^2)\big]\no
&&-\mdel^2 \big[\MOA_{4-5+6} (2s_2-t+2u)-\mnuc \MOA_{1-2+3} (3 u+s_2-4\mnuc^2)\big]\no
&&+\mdel\big[ \mnuc  \MOA_{4-5+6} (\Sigma_{s_2}+s_2+t_1-m_N^2)\no
&&+ \MOA_{1-2+3} \left(\Sigma_{s_2}t_1+\Sigma_{s_2}^\prime\mpir^2\right)\big]
\bigg\} \ .
\eea
\item {Diagram $T^{G\Delta}_{112}$:}  
 \bea
\mathcal{H}^+_\mu=2H^-_\mu&=&-\frac{\h^2}{18 \F \mdel^2 (\mdel^2 - u)(t_1-\mpir^2)}\bigg\{\big[\mnuc \MOA_{1-2+3}+\MOA_{4-5+6} \big] \Sigma_{u}\Sigma_{u}^\prime\no
&&+\mdel^3 \big[-4 \mnuc \MOA_{4-5+6}+\MOA_{1-2+3} (3s_2+u- 4\mnuc^2)\big]\no
&&+\mdel^2 \big[\MOA_{4-5+6} (t-2s_2-2u)+\mnuc \MOA_{1-2+3} (3 s_2+u-4\mnuc^2)\big]\no
&&+\mdel\big[ \mnuc  \MOA_{4-5+6} (\Sigma_{u}+u+t_1-m_N^2)\no 
&&- \MOA_{1-2+3} \left(\Sigma_{u}t_1+\Sigma_{u}^\prime\mpir^2\right)\big]
\bigg\} \ .
\eea

\end{itemize}
\subsection{At \texorpdfstring{$\mathcal{O}(p^{5/2})$:}{O(p5/2):}}
\begin{itemize}
\item {Diagram $T^{A\Delta}_{21}$:}
\bea
\mathcal{H}^+_\mu&=&\frac{\h b_1}{36 \F \mdel^2 (\mdel^2 - s_2)}\bigg\{
2\big[\mdel \mnuc (\Delta_{s_2} + t_1 - 2 \mdel^2) \no
&+& \Sigma_{s_2} (t_1-2 \mnuc^2 ) + \mdel^2 (t - 2 u)\big]\MOV_1
+2 \mdel (\mdel \Dst+\mnuc \Sst)\MOV_2\no
&-&2 \big[(\mdel + \mnuc) (\mdel \mpir^2 + \mnuc \Sst) + (s_2 + \mdel \mnuc) \Sst\big]\MOV_3\no
&+&2\big[\mnuc(\mdel^2-\Sst)+\mdel(\mdel^2-\mpir^2)\big]\MOV_4
+2 \mdel \big[\Sst - 2 \mdel ( \mdel + \mnuc)\big]\MOV_5\no
&-&2 \big[\mdel (2 \mdel^2 + \mpir^2) + \mnuc (2 \mdel^2 + \Sst)\big]\MOV_6+\big[8 \mdel^2 \mnuc (\mdel + \mnuc) \no
&-&  \mdel \mnuc (2 \Sst + \Sstpr) - 
  \Sst \Sstpr +  \mdel^2 (\Sst - 3 t + 2 t_1)\big]\MOV_7\no
 &+&\big[ \mdel^2 (\mdel+\mnuc) (3 \Dst+\Dstpr+3 t-2 t_1)+ \mnuc \Sst \Sstpr\no
 &&-\mdel  \big((\mnuc^2-s_2)^2-( \mpir^2+\Sst) t_1\big)\big]\MOV_8
\bigg\}=-2\mathcal{H}^-_\mu \ .
\eea
\item {Diagram $T^{C\Delta}_{12}$:}
\bea
\mathcal{H}^+_\mu&=&\frac{\h b_1}{36\F \mdel^2 (\mdel^2 - u)}\bigg\{
2 \big[6 \mdel^3 \mnuc + 3 \mdel^2 (2 \mnuc^2 - \mpir^2 + \Su)\big]\MOV_1\no
&+&2 \big[ \mdel^2 \mpir^2 + \mdel \mnuc ( \mpir^2 + 2 \Su) + \Su (\mnuc^2 + u)\big]\MOV_{2-1}\no
&-&2 \mdel (\mnuc \Su + \mdel \Du)\MOV_3+2 \big[ \mdel (\mdel^2 - \mpir^2) + \mnuc ( \mdel^2 - \Su)\big]\MOV_4\no
&+&2 \big[2 \mdel^2 (\mdel + \mnuc) + \mnuc \Su + \mdel \mpir^2\big]\MOV_5+2 \mdel \big[2 \mdel (\mdel + \mnuc) - \Su\big]\MOV_6\no
&-&\big[ \mdel \mnuc ( 8\mdel^2+\mpir^2-t_1-3 \Su)+\mdel^2 (8 \mnuc^2-3 t+2 t_1+\Su)- \Su \Supr\big]\MOV_7\no
&+& \big[\mdel^2 (2 \mnuc^2 - 3 s_2 + u) (\mdel + \mnuc) + \mnuc \Su \Supr \no
&&- 
   \mdel \big((\mnuc^2 - u)^2 - t_1 (\mpir^2 + \Su)\big)\big]\MOV_8
\bigg\}=2\mathcal{H}^-_\mu \ .
\eea

\end{itemize}

\section{Chiral-expansion-suited operators\label{sec:CESbasis}}
As mentioned in Section~\ref{sec:renormalization}, the vector and axial-vector operators given in Eqs.~\eqref{eq:OAs} and~\eqref{eq:OVs} are not suited to perform a chiral expansion. In practice, we prefer to use the following axial-vector operators
\bea\label{eq:CESbasisOA}
\MOACES_{\mu,1}&=&q_\mu\ ,\quad
\MOACES_{\mu,2}=k_\mu\ ,\quad
\MOACES_{\mu,3}=P_\mu \ ,\no
\MOACES_{\mu,4}&=&\frac{1}{2}[\slashed{k},\slashed{q}]\,q_\mu\ ,\quad
\MOACES_{\mu,5}=\frac{1}{2}[\slashed{k},\slashed{q}]\,k_\mu\ ,\quad
\MOACES_{\mu,6}=\frac{1}{2}[\slashed{k},\slashed{q}]\,P_\mu\ ,\no
\MOACES_{\mu,7}&=&\frac{1}{2}[\gamma_\mu,\slashed{q}]\ ,\quad
\MOACES_{\mu,8}=\frac{1}{2}[\gamma_\mu,\slashed{k}]\ ,
\eea
with $P^\mu\equiv{(p^\mu_1+p^\mu_2)}/{2}$. For vector operators, we follow the basis proposed in Ref.~\cite{Adler:1968tw}:
\bea\label{eq:CESbasisOV}
\MOVCES_{\mu,1}&=&\frac{1}{2}\gamma_5[\gamma_\mu,\slashed{k}]\ ,\no
\MOVCES_{\mu,2}&=&2\gamma_5(P_\mu q\cdot k-P\cdot k\, q_\mu)\ ,\no
\MOVCES_{\mu,3}&=&\gamma_5(\gamma_\mu q\cdot k-\slashed{k}\, q_\mu)\ ,\no
\MOVCES_{\mu,4}&=&2\gamma_5\big\{(\gamma_\mu P\cdot k-\slashed{k}\,P_\mu)-\frac{1}{2}m_N[\gamma_\mu,\slashed{k}]\big\}\ ,\no
\MOVCES_{\mu,5}&=&\gamma_5(k_\mu q\cdot k-k^2\,q_\mu)\ ,\no
\MOVCES_{\mu,6}&=&\gamma_5(k_\mu\slashed{k}-k^2\,\gamma_\mu)\ ,
\eea
for which the vector-conservation assumption is automatically implemented.
The axial-vector amplitudes in the new basis can be obtained through
\bea
\Aone &=& A_1+\frac{1}{2}(A_2-A_3)+A_7+\frac{s_2-u}{8 \mnuc}(2 A_4+A_5-A_6) \ ,\no
\Atwo &=&\frac{1}{2}(A_2+A_3)-\frac{s_2-u}{8 \mnuc}(A_5-A_6) \ ,\no
\Athr &=& A_2+A_3+\frac{s_2-u}{4\mnuc}(A_5+A_6)+\frac{1}{\mnuc}A_8\ , \no
\Afou &=& \frac{1}{4 \mnuc}(2 A_4+A_5-A_6)\ ,\no
\Afiv &=& -\frac{1}{4 \mnuc}(A_5-A_6)\ ,\no
\Asix &=& \frac{1}{2 \mnuc}(A_5+A_6)\ ,\no
\Asev &=& A_7+\frac{1}{2 \mnuc}A_8\ ,\no
\Aeig &=& -\frac{1}{2 \mnuc}A_8\ ,
\eea
while the vector amplitudes are
\bea
\Vone &=& \frac{(\mpir^2-t+t_1) }{2 (\mnuc^2-u)}(V_1-2 \mnuc V_4)+\frac{(\mpir^2-t-t_1) }{4 (\mnuc^2-u)}(V_2-V_3)\no
&&+\frac{(s_2-u) }{4 (\mnuc^2-u)}(V_2+V_3)+\frac{\mnuc (t-\mpir^2)}{\mnuc^2-u} V_5-\frac{\mnuc t_1 }{\mnuc^2-u}V_6\ ,\no
\Vtwo &=& \frac{1}{\mnuc^2-u}(2 \mnuc V_4-V_1)+\frac{(\mpir^2-t)(2 \mnuc V_5-V_2)+t_1 (2 \mnuc V_6-V_3)}{(\mpir^2-t+t_1) (\mnuc^2-u)}\ , \no
\Vthr &=& V_4+\frac{1}{2}(V_5-V_6)\ ,\no
\Vfou &=& \frac{1}{2}(V_5+V_6)\ ,\no
\Vfiv &=& \frac{V_3-V_2+2 \mnuc (V_5-V_6)}{\mpir^2-t+t_1}\ ,\no
\Vsix &=& \frac{1}{2}(V_5-V_6)\ .
\eea
In the chiral expansion of $\Delta$-less amplitudes, we treat
\bea
&&\MOACES_{\mu,3}\sim\mathcal{O}(1)\ ,\quad\MOACES_{\mu,1,2,7,8}\sim \MOVCES_{\mu,1,4}\sim \mathcal{O}(p)\ ,\quad
\MOACES_{\mu,6}\sim \MOVCES_{\mu,3,6}\sim \mathcal{O}(p^2)\ ,\no
&&\MOACES_{\mu,4,5}\sim\MOVCES_{\mu,2}\sim  \mathcal{O}(p^3)\ ,\quad 
\MOVCES_{\mu,5}\sim  \mathcal{O}(p^4)\ ,
\eea
and 
\bea
\mnuc\sim \mathcal{O}(1)\ ,\quad s_2-\mnuc^2\sim u-\mnuc^2\sim \mathcal{O}(p)\ ,\quad \mpir^2\sim t_1\sim t\sim\mathcal{O}(p^2)\ .
\eea

\section{Renormalization factors and \texorpdfstring{$\beta$}{beta} functions \label{sec:betas}}
\subsection{Renormalization factors}
The relevant scalar loop functions are defined by
\bea
A_0[m_a^2]&=&\frac{(2\pi\mu^{4-d})}{i\pi^2}\int\frac{{\rm d}^d k}{k^2-m_a^2}\ ,\no
B_0[p^2,m_a^2,m_b^2]&=&\frac{(2\pi\mu^{4-d})}{i\pi^2}\int\frac{{\rm d}^d k}{\big[k^2-m_a^2\big]\big[(k+p)^2-m_b^2\big]}\ ,
\eea
with $\mu$ being the renormalization scale introduced in dimensional regularization.
The explicit form for the one-point one-loop function reads 
\bea
{A}_0[m_a^2]=-m_a^2\left(R+\ln\frac{m_a^2}{\mu^2}\right) ,
\eea
and the scalar two-point one-loop integral has the following analytical form
\bea
{B}_0[p^2,m_a^2,m_b^2]&=&-R+
1-\ln\frac{m_b^2}{\mu^2}+
\frac{m_a^2-m_b^2+p^2}{2\,p^2}\ln\frac{m_b^2}{m_a^2}\no
&& +
\frac{p^2-(m_a-m_b)^2}{p^2}\rho_{ab}(p^2)
\ln\frac{\rho_{ab}(p^2)-1}{\rho_{ab}(p^2)+1}\ ,
\eea
with
\bea
\rho_{ab}(p^2)\equiv\sqrt{\frac{p^2-(m_a+m_b)^2}{p^2-(m_a-m_b)^2}}.
\eea
The UV divergence is contained in the quantity $R={2}/{(d-4)}+\gamma_E-1-\ln(4\pi)$, being $\gamma_E$ the Euler constant. We denote $A_0$ and $B_0$ loop integrals with removed UV-divergent parts (multiples of $R$) by $\bar{A}_0$ and $\bar{B}_0$, respectively.

To proceed, the nucleon and pion wave function renormalization constants can be written as
\bea
 {\cal Z}_N=1+\delta_{\mathcal{Z}_N}^{(2)}\ ,\quad {\cal Z}_\pi=1+\delta_{\mathcal{Z}_\pi}^{(2)}\ ,
\eea
respectively, where the $\mathcal{O}(p^2)$ parts are
\bea
\delta_{\mathcal{Z}_N}^{(2)}&=&-\frac{3 \g^2 }{64\pi^2 \F^2 (\mpir^2-4 \mnuc^2)}\bigg\{
(12 \mnuc^2-5 \mpir^2) \LFAo[\mpir^2]+4 \mpir^2 (-\mnuc^2+\LFAo[\mnuc^2]\no
&+&(\mpir^2-3 \mnuc^2) \LFBo[\mnuc^2,\mpir^2,\mnuc^2])\bigg\}\ ,\no
\delta_{\mathcal{Z}_\pi}^{(2)}&=&-\frac{2}{3\F^2} \bigg\{3 l_4 \mpir^2+\frac{\LFAo[\mpir^2]}{16\pi^2}\bigg\}\ .
\eea

The relations between the renormalized (or chiral limit) masses and the physical ones read
\bea\label{eq:renfactors1}
m_N=\tilde{m}-4\tilde{c}_1\mpir^2+\delta_{m_N}^{(3)}\ ,\quad M_\pi^2=M^2(1+\delta_{M_\pi^2}^{(2)})\ ,\no
\eea
with
\bea
\delta_{M_\pi^2}^{(2)}&=&\frac{2 l_3^r \mpir^2}{\F^2}-\frac{\LFAoBar[\mpir^2]}{32\pi^2 \F^2}\ ,\no
\delta_{m_N}^{(3)}&=&\frac{3 \g^2 \mnuc \mpir^2}{32\pi^2 \F^2}\bigg\{ \LFBoBar[\mnuc^2,\mpir^2,\mnuc^2]-\bigg(1+\frac{\LFAoBar[\mnuc^2]}{\mnuc^2}\bigg)\bigg\}\ .
\eea
Likewise, for the leading couplings $g_A$ and $F_\pi$, one has
\bea\label{eq:renfactors2}
&&g_A=\tilde{g}\left(1+\frac{4d_{16}^r\mpir^2}{g_A}+\delta_{g_A}^{(2)}\right)\ ,\quad F_\pi=F(1+\delta_{F_\pi}^{(2)})\ ,
\eea
with
\bea
\delta_{F_\pi}^{(2)}&=&\frac{l_4^r \mpir^2}{\F^2}+\frac{\LFAoBar[\mpir^2]}{16\pi^2\F^2}\ ,\no
\delta_{g_A}^{(2)}&=&\frac{1}{16\pi^2\F^2 (\mpir^2-4 \mnuc^2)}\bigg\{
\big[(1+4 \g^2) \mpir^2-4 (1+2 \g^2) \mnuc^2\big] \LFAoBar[\mpir^2]+\mpir^2 \big[4 \g^2 \mnuc^2\no
&&-4 \g^2 \LFAoBar[\mnuc^2]+\big(8 (1+\g^2) \mnuc^2-(2+3 \g^2) \mpir^2\big) \LFBoBar[\mnuc^2,\mpir^2,\mnuc^2]\big]
\bigg\} \ .
\eea

\subsection{UV-\texorpdfstring{$\beta$}{beta} functions}
In Eq.~\eqref{eq:UVshift}, the $\beta$-functions corresponding to the infinite parts of counter terms for the pionic LECs $l_i$ ($i=3,4,6$) are
\bea\label{eq:betaell}
\beta_{l_3} =-\frac{1}{4}\ ,\quad
\beta_{l_4} =1\ ,\quad
\beta_{l_6} =-\frac{1}{6}\ .
\eea
For the constants appearing in the LO $\pi N$ Lagrangian, we get
\bea
\beta_{m}=\frac{3 {\gA}^2 {\mn}^3}{2 F^2}\ , \quad
\beta_{g}=\frac{\gA (-2 + \gA^2) \mn^2}{F^2}\ .
\eea
The ones for $c_j$ read
\bea
\beta_{c_1}&=&-\frac{3 \gA^2 \mn}{8 F^2}\ ,\quad
\beta_{c_2}=\frac{(-1+\gA^2)^2 \mn}{2 F^2}\ ,\no
\beta_{c_3}&=&\frac{(1-6 \gA^2+\gA^4) \mn}{4 F^2}\ ,\quad
\beta_{c_4}=\frac{(-1-2 \gA^2+3 \gA^4) \mn}{4 F^2}\ ,\quad
\beta_{c_6}=0\ ,
\eea
and the ones for $d_k$ are given by
\bea
4\beta_{d_1}&=&\beta_{d_{14}}=\frac{1}{2}\beta_{d_{15}}=\beta_{d_{23}}=-\frac{(-1+\gA^2)^2}{4 F^2}\ ,\no
\beta_{d_2}&=&\frac{2-5 \gA^2+3 \gA^4}{24 F^2}\ ,\nonumber\\
\beta_{d_3}&=&\beta_{d_8}=\beta_{d_{18}}=\beta_{d_{20}}=\beta_{d_{21}}=\beta_{d_{22}}=0\ ,\nonumber\\
4\beta_{d_5}&=&-\beta_{d_6}=-\frac{1}{3g}\beta_{d_{16}}=-\frac{-1+\gA^2}{12 F^2}\ .\nonumber\\
\eea

\subsection{EOMS-\texorpdfstring{$\tilde{\beta}$}{beta} functions}
In Eq.~\eqref{eq:finiteshift}, the EOMS-$\bar{\beta}$ functions are responsible for the finite shifts of the LECs, which as a result cancel the PCB terms from loops. Only for the LECs in $\mathcal{L}_{\pi N}^{(1)}$ and $\mathcal{L}_{\pi N}^{(2)}$ one needs to carry out finite shifts. For the LO pion-nucleon parameters, the $\bar{\beta}$ functions are
\bea
\tilde{\beta}_{g_A} &=& \gA^3 \mn+\frac{\gA (2-\gA^2)}{\mn} \LFAoBar[\mn^2]\ ,\no
\tilde{\beta}_{m} &=& -\frac{3}{2} \gA^2 \LFAoBar[\mn^2]\ ,\no
\eea
while the NLO ones read
\bea
\tilde{\beta}_{c_1}&=&\frac{3 }{8}\gA^2+\frac{3\,\gA^2 }{8\, \mn^2} \LFAoBar[\mn^2]\ ,\no
\tilde{\beta}_{c_2}&=&-\frac{2+\gA^4}{2}-\frac{(\gA^2-1)^2}{2 \mn^2} \LFAoBar[\mn^2]\ ,\no
\tilde{\beta}_{c_3}&=&\frac{9}{4} \gA^4-\frac{1-6 \gA^2+\gA^4}{4 \mn^2} \LFAoBar[\mn^2]\ ,\no
\tilde{\beta}_{c_4}&=&-\frac{1}{4} \gA^2 (5+\gA^2)+\frac{1+2 \gA^2-3 \gA^4}{4 \mn^2} \LFAoBar[\mn^2] \ ,\no
\tilde{\beta}_{c_6}&=& 0\ .
\eea

\end{widetext}

\section{Isobaric-frame amplitudes\label{sec:isobaramp}}
Following Ref.~\cite{Adler:1968tw}, the multipole expansion of the scattering matrix element is performed in the isobaric frame. The linear transformations expressing $\mathscr{F}_i^V$ and $\mathscr{G}_i^A$, defined in Eq.~\eqref{eq:isobaricamp}, in terms of $V_i$ and $A_i$, defined in Eq.~\eqref{eq:lordecom}, are given below. For the vector amplitudes, they are:
\bea
\mathscr{F}_1^V&=&-\frac{N_1}{N_2}\bigg\{\qdotq V_7+N_2^2(q_0V_7-V_8)\bigg\}\ ,\nonumber\\
\mathscr{F}_2^V&=&\frac{\absk\absq}{N_1N_2}\bigg\{(N_2^2+q_0)V_7+V_8\bigg\}\ ,\nonumber\\
\mathscr{F}_3^V&=&-\frac{\absk\absq}{N_1N_2}\bigg\{\qdotq (V_4 - V_6) + N_2^2 (V_1 - V_3 + q_0 V_4 - q_0 V_6 + 2 V_7)\bigg\}\ ,\nonumber\\
\mathscr{F}_4^V&=&-\frac{\qdotq N_1}{N_2}\bigg\{V_3 - V_1 + (N_2^2 + q_0) (V_4 - V_6) - 2 V_7\bigg\}\ ,\nonumber\\
\mathscr{F}_5^V&=&-\frac{1}{N_1N_2t_1}\bigg\{k_0 \big[\kdotk \big(\qdotq V_5 + N_2^2 (V_2 + q_0 V_5)\big) + 
     \kdotq \big(\qdotq (-V_4 + V_6) \nonumber\\
     &&+ 
        N_2^2 (-V_1 + V_3 - q_0 V_4 + q_0 V_6 - 2 V_7)\big) - 
     N_1^2 \big(\qdotq V_7 + N_2^2 (q_0 V_7 - V_8)\big)\big]  \nonumber\\
     &&+
  \absk^2 \big[\qdotq (q_0 V_4 + p_{10} V_5 + p_{20} V_6 - V_7) + 
     N_2^2 (p_{10} V_2 + p_{20} V_3 \nonumber\\
     &&+ q_0 (V_1 + q_0 V_4 + p_{10} V_5 + p_{20} V_6 + V_7) + 
        V_8)\big]\bigg\}\ ,\nonumber\\
\mathscr{F}_6^V&=&\frac{\absq}{\absk N_1N_2t_1}\bigg\{k_0 \kdotq N_1^2 \big[-V_1 + V_3 + (N_2^2 + q_0) (V_4 - V_6) - 2 V_7\big] \nonumber\\
&&- \absk^2 \big[k_0 \big((N_2^2 + q_0) V_7 + V_8\big) + 
    N_1^2 \big((-k_0 - p_{10}) V_2 - p_{20} V_3 + q_0^2 V_4 \nonumber\\
   &&+    q_0 (-V_1 + N_2^2 V_4 + (k_0 + p_{10}) V_5 + p_{20} V_6 - V_7) \nonumber\\
   &&+ 
       N_2^2 ((k_0 + p_{10}) V_5 + p_{20} V_6 + V_7) + V_8\big)\big]\bigg\}\ ,
\eea
with the normalization factors $N_1=\sqrt{p_{10}+m_N}$ and $N_2=\sqrt{p_{20}+m_N}$. Likewise,
for the axial-vector amplitudes, we have
\bea
\mathscr{G}_1^A&=&\frac{\absq}{N_1N_2}\bigg\{ A_8 N_1^2 + A_7 \big[2 \kdotq + N_1^2 (N_2^2 + q_0)\big]\bigg\}\ ,\nonumber\\
\mathscr{G}_2^A&=&-\frac{\absk}{N_1N_2}\bigg\{A_7 (N_2^2 q_0 + \qdotq)-A_8 N_2^2\bigg\}\ ,\nonumber\\
\mathscr{G}_3^A&=&-\frac{\absk\absq^2}{N_1N_2}\bigg\{A_3-A_1+(A_4-A_6)(N_2^2+q_0)\bigg\}\ ,\nonumber\\
\mathscr{G}_4^A&=&-\frac{\absq N_1}{N_2}\bigg\{N_2^2 (A_1 - A_3 + 2 A_7 + A_4 q_0 - A_6 q_0) + (A_4 - A_6) \qdotq\bigg\}\ ,\nonumber\\
\mathscr{G}_5^A&=&\frac{\absq}{\absk N_1N_2}\bigg\{\absk^2 k_0 \bigg[-A_2 + A_5 (N_2^2 + q_0)\big] - 
  k_0 \kdotq \big[-A_1 + A_3\nonumber\\
  && + (A_4 - A_6) (N_2^2 + q_0)\big] + 
  \absk^2 \big[A_8 - A_2 p_{10} - A_3 p_{20} + A_7 (N_2^2 - q_0)\nonumber\\
  && - 
     A_1 q_0 + (N_2^2 + q_0) (A_5 p_{10} + A_6 p_{20} + A_4 q_0)\big] \nonumber\\
     &&+ 
  k_0 \big[A_8 N_1^2 + A_7 \big(2 \kdotq + N_1^2 (N_2^2 + q_0)\big)\big]\bigg\}\ ,\nonumber\\
  \mathscr{G}_6^A&=&\frac{1}{\absk^2N_1N_2}\bigg\{A_8 \absk^2 (k_0 + N_1^2) N_2^2 + A_2 \absk^2 N_1^2 N_2^2 (k_0 + p_{10})  \nonumber\\
 &&+ N_2^2 \big[N_1^2 \big((A_3-A_1  - 2 A_7) k_0 \kdotq + 
        A_3 \absk^2 p_{20}\big) + \big(A_7 \absk^2 (N_1^2-k_0 ) \nonumber\\
        &&+ 
        N_1^2 (A_1 \absk^2 - A_4 k_0 \kdotq + A_6 k_0 \kdotq + 
           A_5 \absk^2 (k_0 + p_{10}) + A_6 \absk^2 p_{20})\big) q_0 \nonumber\\
           &&+ 
     A_4 \absk^2 N_1^2 q_0^2 \big]+ (-A_7 \absk^2 (k_0 + N_1^2) + 
     N_1^2 (-A_4 k_0 \kdotq + A_6 k_0 \kdotq \nonumber\\
     &&+ A_5 \absk^2 (k_0 + p_{10}) + 
        A_6 \absk^2 p_{20} + A_4 \absk^2 q_0)) \qdotq\bigg\}\ ,\nonumber\\
\mathscr{G}_7^A&=& \frac{\absk\absq}{N_1N_2}\bigg\{ A_8 - A_2 p_{10} - A_3 p_{20} + A_7 (N_2^2 - q_0) - 
  A_1 q_0 
  \nonumber\\&&
  + (N_2^2 + q_0) (A_5 p_{10} + A_6 p_{20} + A_4 q_0)\bigg\} \ ,\nonumber\\
  \mathscr{G}_8^A&=&\frac{N_1}{N_2}\bigg\{N_2^2 \big[A_8 + A_2 p_{10} + A_3 p_{20} + 
     q_0 (A_1 + A_7 + A_5 p_{10} + A_6 p_{20} + A_4 q_0)\big]
     \nonumber\\&&
   + (-A_7 + A_5 p_{10} + 
     A_6 p_{20} + A_4 q_0) \qdotq\bigg\} \ .
\eea
The above expressions are deduced in the CM of the outgoing pion-nucleon pair. Therefore the energies and momenta can be written as functions of $W=\sqrt{s_2}$ and $t_1$:
\bea
&&p_{10} = \frac{W^2+m_N^2-t_1}{2W}\ ,\quad k_0 =\frac{W^2-m_N^2+t_1}{2W}\ , \quad \absk =\sqrt{k_0^2-t_1}\ ,\nonumber\\
&&p_{20} = \frac{W^2+m_N^2-M_\pi^2}{2W}\ ,\quad q_{0} = \frac{W^2-m_N^2+M_\pi^2}{2W} \ ,\quad \absq=\sqrt{q_0^2-M_\pi^2}\ .\nonumber\\
&&
\eea

\bibliography{weak}

\end{document}